%% file: hyb.tex
\documentclass[prd,onecolumn,showkeys,
preprint,
nofootinbib,
 amsmath,amssymb,
 aps,
]{revtex4-2}
\usepackage{yfonts}
\usepackage{graphicx}% Include figure files
\usepackage{dcolumn}% Align table columns on decimal point
\usepackage{bm}% bold math
\usepackage{nicefrac}
\usepackage{color}
\usepackage{xcolor}
\definecolor{darkblue}{RGB}{0,0,196}
\definecolor{darkgreen}{RGB}{0,120,0}
\usepackage[colorlinks=true,linktocpage=true,linkcolor=darkblue,citecolor=red,urlcolor=darkblue]{hyperref}
\usepackage{hyperref}
\usepackage{appendix}
\usepackage{eqnarray}
\usepackage{ccaption}

\newcommand{\bea}{\begin{eqnarray}}
\newcommand{\eea}{\end{eqnarray}}
\newcommand{\bel}[1]{\begin{eqnarray}\label{#1}}
\newcommand{\eel}{\end{eqnarray}}
\newcommand{\beq}{\begin{equation}}
\newcommand{\eeq}{\end{equation}}

\def\LB{\left(}
\def\RB{\right)}
\def\LSB{\left[}
\def\RSB{\right]}
\def\LAB{\langle}
\def\RAB{\rangle}

\newcommand{\nn}{\nonumber}

\newcommand{\EQ}[1]{Eq.~(\ref{#1})}
\newcommand{\EQn}[1]{(\ref{#1})}

\newcommand{\EQSTWO}[2]{Eqs.~(\ref{#1})~and~(\ref{#2})}
\newcommand{\EQSTWOn}[2]{(\ref{#1})~and~(\ref{#2})}

\newcommand{\EQSM}[2]{Eqs.~(\ref{#1})--(\ref{#2})}
\newcommand{\EQSMn}[2]{(\ref{#1})--(\ref{#2})}

\newcommand{\SEC}[1]{Sec.~\ref{#1}}

\newcommand{\CIT}[1]{Ref.~\citep{#1}} 
\newcommand{\CITS}[1]{Refs.~\citep{#1}} 
\newcommand{\CITn}[1]{\citep{#1}} 

\newcommand{\p}{\partial}
\newcommand{\dd}{\mathrm{d}}

\def\epsUabgd{\epsilon^{\alpha \beta \gamma \delta}}

\newcommand{\av}{{\boldsymbol a}} 
\newcommand{\bv}{{\boldsymbol b}} 
 
\newcommand{\kv}{{\boldsymbol k}}

\newcommand{\pv}{{\boldsymbol p}}
\newcommand{\tv}{{\boldsymbol t}}

\newcommand{\sv}{{\boldsymbol s}}

\newcommand\omv{{\boldsymbol \omega}}

\newcommand{\f}[2]{\frac{#1}{#2}}
\newcommand{\onehalf}{{\nicefrac{1}{2}}} 
 
\newcommand{\threefourths}{{\nicefrac{3}{4}}} 

\def\spin{\,\textgoth{s:}}
\def\spinl{|{\boldsymbol s}_*|}

\begin{document}

%\preprint{APS/123-QED}

\title{Hybrid approach to perfect and dissipative spin hydrodynamics}
% Force line breaks with \\
%\thanks{A footnote to the article title}%

\author{Zbigniew Drogosz}
\email{zbigniew.drogosz@alumni.uj.edu.pl}
\affiliation{Institute of Theoretical Physics, Jagiellonian University, PL-30-348 Krak\'ow, Poland}

\author{Wojciech Florkowski}
\email{wojciech.florkowski@uj.edu.pl}
\affiliation{Institute of Theoretical Physics, Jagiellonian University, PL-30-348 Krak\'ow, Poland}

\author{Mykhailo Hontarenko}
\email{mykhailo.hontarenko@student.uj.edu.pl}
\affiliation{Institute of Theoretical Physics, Jagiellonian University, PL-30-348 Krak\'ow, Poland}

\date{\today}% It is always \today, today,
             %  but any date may be explicitly specified

\begin{abstract}
A hybrid framework of spin hydrodynamics is proposed that combines the results of kinetic theory for particles with spin 1/2 with the Israel-Stewart method of introducing nonequilibrium dynamics. The framework of kinetic theory is used to define the perfect-fluid description that conserves baryon number, energy, linear momentum and spin part of angular momentum. This leads to the entropy conservation although, in the presence of spin degrees of freedom, the perfect-fluid formalism includes extra terms whose structure is usually attributed to dissipation. The genuine dissipative terms appear from the condition of positive entropy production in nonequilibrium processes. They are responsible for the transfer between the spin and orbital parts of angular momentum, with the total angular momentum being conserved. 
\end{abstract}


%Use showkeys class option if keyword
\maketitle

%\tableofcontents

%%%%%%%%%%%%%%%%%%%%%%%%%%%%%%%%%%%%%%%%%%%%%%
\section{Introduction}

New results regarding the spin polarization of particles produced in relativistic collisions of heavy ions \cite{STAR:2017ckg, STAR:2018gyt, STAR:2019erd,ALICE:2019aid} opened new research perspectives. In addition to standard measurements of particle abundances, the momentum distributions and various correlations, it is now possible to study completely new observables related to spin degrees of freedom. For a recent review see, for example, Ref.~\cite{Niida:2024ntm}. 

Taking theoretical description of heavy ion collisions as a reference point, it becomes a very important issue to take into account the spin degrees of freedom in the formalism of relativistic hydrodynamics~\CITn{
%%%%%%%%%%%%%%%%%%%%%%%%%%%%%%
Becattini:2009wh,  Becattini:2011zz, Becattini:2013fla, Becattini:2020qol, Becattini:2021iol, Palermo:2024tza, 
%%%%%%%%%%%%%%%%%%%%%%%%%%%%%%
Florkowski:2017ruc, Florkowski:2017dyn, Bhadury:2020puc, Bhadury:2020cop, Bhadury:2022ulr, Singh:2022ltu,
Weickgenannt:2019dks, Weickgenannt:2021cuo, Weickgenannt:2020aaf, Wagner:2022amr, Weickgenannt:2022zxs, Weickgenannt:2023nge, Wagner:2024fhf,
Hu:2021pwh, 
Li:2020eon,  Shi:2020htn, 
%%%%%%%%%%%%%%%%%%%%%%%%%%%%%%
Hattori:2019lfp, Fukushima:2020ucl, Daher:2022xon, Daher:2022wzf, Sarwar:2022yzs, Wang:2021ngp, Biswas:2022bht, Biswas:2023qsw, Xie:2023gbo, Daher:2024ixz, Xie:2023gbo, Ren:2024pur, Daher:2024bah, 
Gallegos:2021bzp, Hongo:2021ona, 
Kumar:2023ojl, She:2021lhe, 
%%%%%%%%%%%%%%%%%%%%%%%%%%%%%%
Montenegro:2017rbu, Montenegro:2020paq, Goncalves:2021ziy}. The use of the latter to describe expansion of the produced matter has become the main paradigm used in the so-called standard model of heavy-ion collisions~\CITn{Florkowski:2010zz,Gale:2013da,Jaiswal:2016hex}. 

Despite significant progress achieved in the development of the idea of spin hydrodynamics, there exists an important problem of parallel and independent elaboration of different concepts and their often unclear relation to each other. The construction of the formalism of spin hydrodynamics currently pursues the following main paths: {\bf (i)} only the gradients of hydrodynamic fields on the freezeout hypersurface are used to determine the final hadron polarization~\CITn{Becattini:2009wh,  Becattini:2011zz, Becattini:2013fla, Palermo:2024tza, Becattini:2021iol}, {\bf (ii)} the kinetic theory of particles with spin $\onehalf$~\CITn{Florkowski:2017ruc, Florkowski:2017dyn, Bhadury:2020puc, Bhadury:2020cop, Bhadury:2022ulr, Singh:2022ltu, Weickgenannt:2019dks, Weickgenannt:2021cuo, Weickgenannt:2020aaf, Wagner:2022amr, Weickgenannt:2022zxs, Weickgenannt:2023nge, Wagner:2024fhf, Hu:2021pwh, Li:2020eon,  Shi:2020htn} is taken as the starting point and subsequently the hydrodynamic equations are obtained from the moments of kinetic equations (in this case one may start from the quantum field theory and subsequently apply a semiclassical expansion, or one uses the classical approach to spin; many works also combine these two approaches, whose mutual relation is discussed, for example, in~\CIT{Florkowski:2018fap}), {\bf (iii)}~the formalism is constructed by referring to mathematically permissible expressions for the energy-momentum and spin tensors, and the conservation laws along with the law of entropy increase are applied~\CITn{Hattori:2019lfp, Fukushima:2020ucl, Daher:2022xon, Daher:2022wzf, Sarwar:2022yzs, Wang:2021ngp, Biswas:2022bht, Biswas:2023qsw, Xie:2023gbo, Daher:2024ixz, Xie:2023gbo, Ren:2024pur, Daher:2024bah, Gallegos:2021bzp, Hongo:2021ona, Kumar:2023ojl, She:2021lhe}, and {\bf (iv)} the spin-extended Lagrangian formalism is used~\CITn{Montenegro:2017rbu,Montenegro:2020paq,Goncalves:2021ziy}.

In this work we indicate how one can consistently combine some of the results obtained in the areas {\bf (ii)} and {\bf (iii)}. This also leads us to take a new look at the role played by the thermal vorticity and gradient expansion presently used in spin hydrodynamics. 

Many formal developments of spin hydrodynamics in {\bf (iii)} closely follow the original ideas of Israel and Stewart (IS)~\CITn{Israel:1979wp}. The main difficulty in the application of these ideas to spin-polarizable systems is inconsistent treatment of the expansion in the spin polarization tensor $\omega^{\mu\nu}$. One assumes that the spin density tensor $S^{\mu\nu}$ is of the zeroth order in $\omega^{\mu\nu}$ although in the leading order one finds that $S^{\mu\nu}$ is proportional to $\omega^{\mu\nu}$. This ansatz allows to keep the terms $S^{\mu\nu} \omega_{\mu\nu}$ in thermodynamic relations involving spin—otherwise they should be ignored as quadratic corrections. Another problem of this approach is the use of the spin tensor whose form is inconsistent with the kinetic-theory result. Some of these issues have been clarified in a recent paper on generalized thermodynamic relations~\CITn{Florkowski:2024bfw}. By consistent inclusion of the second-order terms in  $\omega^{\mu\nu}$, one recovers consistent and nontrivial thermodynamic relations. Moreover, an agreement with the kinetic theory is achieved. 

On the kinetic-theory side presented by {\bf (ii)}, many works are restricted to the case of local collisions that prohibit processes leading to the conversion of the orbital part of the angular momentum, $L$, into the spin part, $S$, and the other way round. A solution to this problem proposed in a series of significant papers is to include nonlocal collisions~\CITn{Weickgenannt:2019dks, Weickgenannt:2021cuo, Weickgenannt:2020aaf, Wagner:2022amr, Weickgenannt:2022zxs}. The resulting formalism is, however, very complicated, and it is not clear how nonlocal effects in collisions may affect a causal framework of relativistic hydrodynamics.

In this work we suggest that the formalism of spin hydrodynamics can be developed by combining the most attractive features of the approaches {\bf (ii)} and {\bf (iii)}. At the level of perfect fluid, we propose to rely on the local kinetic theory. However, adding dissipative effects can be done using the IS methods, instead of switching to the formalism with nonlocal collisions. In the perfect-fluid description, one conserves the baryon number, energy, linear momentum and the spin part of the total angular momentum. These conservation laws lead, in a non trivial way, to the conserved entropy current, as shown in~\CITn{Florkowski:2024bfw}. Inclusion of the dissipation with the IS method allows for transfer between $L$ and $S$. 

%\smallskip
The paper is organized as follows: In \SEC{sec:trel} we review thermodynamic relations used in the phenomenological formulation of spin hydrodynamics. In \SEC{sec:kint}, using the kinetic-theory approach, we argue that the naive form of thermodynamic relations should be replaced by the generalized identities, as recently shown in~\CIT{Florkowski:2024bfw} for the Boltzmann statistics. Herein, we demonstrate that the generalized form of thermodynamic relations is valid also in the case of the Fermi-Dirac (FD) statistics. Section~\ref{sec:smallomega} introduces an expansion in the spin polarization tensor that is dimensionless in natural units and whose components become natural expansion parameters. In \SEC{sec:Boltzmann} we restrict our considerations to the case of Boltzmann statistics and explicitly construct all the macroscopic currents (a rather technical further discussion of the FD case is continued in Appendix~\ref{sec:Ztensors}). 

Sections~\ref{sec:trel}--\ref{sec:Boltzmann} may be considered as an extension of~\CIT{Florkowski:2024bfw}. They define a framework of perfect spin hydrodynamics. In \SEC{sec:closetoeq} we show how this approach can be generalized to include dissipation. This is achieved with the Israel-Stewart method based on the analysis of the off-equilibrium entropy production. Although the IS approach is used, in this work we consider only linear terms in gradients, i.e., we remain at the level of the relativistic Navier-Stokes theory (a rather straightforward development of the second-order theory in gradients is left for a separate study). We summarize and conclude in \SEC{sec:summary}. Four Appendices include discussion of certain integrals, the spin integration measure, the pressure of the Fermi-Dirac gas, and the contraction of rank-3 tensors that are antisymmetric in the last two indices.

Before we turn to discussion of physical issues, let us define our notation. For the Levi-Civita tensor $\epsilon^{\mu\nu\alpha\beta}$ we follow the convention $\epsilon^{0123} =-\epsilon_{0123} = +1$. The metric tensor is of the form $g_{\mu\nu} = \textrm{diag}(+1,-1,-1,-1)$. The scalar products for both three- and four-vectors are denoted by a dot, i.e., $a \cdot b = a^0 b^0 - \av \cdot \bv$. Throughout the text we make use of natural units, $\hbar = c = k_B = 1$.

%%%%%%%%%%%%%%%%%%%%%%%%%%%%%%%%%%%%%%%%%%%%%%
\section{Thermodynamic relations}
\label{sec:trel}

In this section, we review the relativistic forms of thermodynamic relations used in relativistic hydrodynamics. We start with the standard case without spin degrees of freedom and then switch to a popular form used in many formulations of spin hydrodynamics. The latter approach is often called the phenomenological version of spin thermodynamic relations. In the next sections, we demonstrate that it must be modified to be consistent with the results of microscopic calculations obtained within the kinetic theory.

%%%%%%%%%%%%%%%%%%%%%%%%%%%%%%%%%%%%%%%%%%%%%%
\subsection{Spinless case}

The standard thermodynamic relations used in relativistic hydrodynamics include the identity
\bel{eq:ext_S}
\varepsilon + P  = T \sigma + \mu n
\eel 
and the first law of thermodynamics
\bel{eq:firstlaw_S}
\dd \varepsilon =T \dd \sigma + \mu \dd n .
\eel 
Here the letters $\varepsilon$, $P$, $T$, $\sigma$, $\mu$ and $n$ denote the local energy density, pressure, temperature, entropy density, baryon chemical potential, and baryon number density, respectively. Equation \EQn{eq:ext_S} follows from  the extensivity of energy, entropy and baryon number. Equations \EQSTWOn{eq:ext_S}{eq:firstlaw_S} imply the Gibbs-Duhem relation
\bel{eq:GD_S}
\dd P = \sigma \dd T + n \dd \mu .  
\eel 
Following the seminal works of Israel and Stewart, one often rewrites \EQSM{eq:ext_S}{eq:firstlaw_S} in a tensor (four-vector) form, which is convenient for further incorporation of dissipative phenomena. This is done by multiplication of \EQSM{eq:ext_S}{eq:GD_S} by the local four-velocity of the fluid $u^\mu$, which leads to the following expressions
\bel{eq:ext_T}
S^\mu_{\rm eq} = \sigma u^\mu = P \beta^\mu - \xi N^\mu_{\rm eq} + \beta_\lambda T^{\lambda\mu}_{\rm eq},
\eel
\bel{eq:firstlaw_T}
\dd S^\mu_{\rm eq} = - \xi \dd N^\mu_{\rm eq} + \beta_\lambda \dd T^{\lambda\mu}_{\rm eq},
\eel
and
\bel{eq:GD_T}
\dd (P\beta^\mu) = N^\mu_{\rm eq} \dd \xi - T^{\lambda\mu}_{\rm eq} \dd \beta_\lambda.
\eel
Here we have introduced a common notation,
\bel{eq:beta}
\beta^\mu = \f{u^\mu}{T}, \quad \beta = \sqrt{\beta^\lambda \beta_\lambda} = \f{1}{T}, \quad \xi = \f{\mu}{T}.
\eel
The four-vector $S^\mu_{\rm eq}$ represents the entropy current, while $N^\mu_{\rm eq}$ and $T^{\lambda\mu}_{\rm eq}$ are the baryon current and energy-momentum tensors for a perfect fluid
\bel{eq:NmuPF}
 N^\mu_{\rm eq} = n u^\mu
\eel
and
\bel{eq:TmunuPF}
T^{\lambda\mu}_{\rm eq} = (\varepsilon+P) u^\lambda u^\mu - P g^{\lambda \mu} = \varepsilon u^\lambda u^\mu - P \Delta^{\lambda \mu},
\eel
where the tensor
\bel{eq:Delta}
\Delta^{\lambda \mu} = g^{\lambda \mu}-u^\lambda u^\mu
\eel%
projects on the space orthogonal to flow. Below, we also introduce the projector $\Delta^{\mu\nu}_{\alpha\beta}$ defined by the expression 
\bel{eq:Delta2}
\Delta^{\mu\nu}_{\alpha\beta} = \f{1}{2} \LB
\Delta^{\mu}_{\alpha} \Delta^{\nu}_{\beta} +
\Delta^{\nu}_{\alpha} \Delta^{\mu}_{\beta} - \f{2}{3} 
\Delta^{\mu\nu} \Delta_{\alpha\beta} \RB.
\eel
The contraction $\Delta^{\mu\nu}_{\alpha\beta} A^{\alpha \beta} \equiv  A^{\LAB \alpha \beta \RAB}$ picks up the orthogonal, symmetric, and traceless part of the tensor $A^{\alpha \beta}$. We also introduce the standard notation where round (squared) brackets denote symmetrization (antisymmetrization) of the Lorentz indices:  $A^{(\mu\nu)}= \frac{1}{2}(A^{\mu\nu}+A^{\nu\mu})$ and $A^{[\mu\nu]}=  \frac{1}{2} (A^{\mu\nu}-A^{\nu\mu})$. Moreover, we introduce an orthogonal projection $T^{\alpha \beta \LAB \gamma \RAB \delta ...} = \Delta^\gamma_\rho T^{\alpha \beta \rho \delta ...}$. If the symbols $\LAB \RAB$ are used to the right of the differential symbol, the contractions with $\Delta$'s should be done after the derivative is calculated first. The transverse gradient $\nabla^\mu$ is defined by contraction $\Delta^{\mu \nu} \p_\nu$.

%%%%%%%%%%%%%%%%%%%%%%%%%%%%%%%%%%%%%%%%%%%%%%
\subsection{Including spin degrees of freedom—phenomenological version}

Many approaches to relativistic spin hydrodynamics extend \EQSM{eq:ext_S}{eq:GD_S} to include the tensor spin chemical potential $\Omega_{\alpha\beta}$ and the spin density tensor $S^{\alpha \beta}$ (which are both rank-2 antisymmetric tensors). They read
\bel{eq:ext_Ss}
\varepsilon + P  = T \sigma + \mu n + \frac{1}{2}
\Omega_{\alpha\beta} S^{\alpha \beta},
\eel 
\bel{eq:firstlaw_Ss}
\dd \varepsilon =T \dd \sigma + \mu \dd n + \frac{1}{2}
\Omega_{\alpha\beta} \dd S^{\alpha \beta},
\eel 
\bel{eq:GD_Ss}
\dd P = \sigma \dd T + n \dd \mu + \frac{1}{2}
S^{\alpha \beta} \dd \Omega_{\alpha\beta}.  
\eel 
By multiplying \EQSM{eq:ext_Ss}{eq:GD_Ss} by the hydrodynamic flow vector $u^\mu$, we obtain
\bel{eq:ext_Ts}
S^\mu_{\rm eq} =  P \beta^\mu - \xi N^\mu_{\rm eq} + \beta_\lambda T^{\lambda\mu}_{\rm eq}
- \frac{1}{2}
\omega_{\alpha\beta} S^{\mu, \alpha \beta}_{\rm eq},
\eel
\bel{eq:firstlaw_Ts}
\dd S^\mu_{\rm eq} = - \xi \dd N^\mu_{\rm eq} + \beta_\lambda \dd T^{\lambda\mu}_{\rm eq}
- \frac{1}{2}
\omega_{\alpha\beta} \dd S^{\mu, \alpha \beta}_{\rm eq},
\eel
\bel{eq:GD_Ts}
\dd (P\beta^\mu) = N^\mu_{\rm eq} \dd \xi - T^{\lambda\mu}_{\rm eq} \dd \beta_\lambda + \frac{1}{2}
S^{\mu, \alpha \beta}_{\rm eq} \dd \omega_{\alpha\beta}.
\eel
The spin polarization tensor that appears here is defined as the ratio
\bel{eq:omega}
\omega_{\alpha\beta} = \f{\Omega_{\alpha\beta}}{T}.
\eel 
Below we use the parametrization introduced in~\CITn{Florkowski:2017ruc}
\bel{eq:ko}
\omega_{\alpha\beta} = k_\alpha u_\beta - k_\beta u_\alpha + t_{\alpha\beta},
\eel 
where 
\bel{eq:tab}
t_{\alpha\beta}=\epsilon_{\alpha\beta\gamma\delta} u^\gamma \omega^\delta,
\eel
and the four-vectors $k$ and $\omega$ are orthogonal to the flow vector $u$ (i.e., $k\cdot u = 0$ and $\omega \cdot u = 0$). One can easily check the following property~\footnote{We use a colon to denote contraction of two rank-2 tensors, e.g., $A:B = A_{\alpha\beta} B^{\alpha\beta}$. Note the order of the indices.}
\bel{eq:ome:om}
\omega:\omega \equiv \omega^{\alpha\beta} \omega_{\alpha \beta} = 2 (k^2 - \omega^2).  
\eel 
Below, we also frequently use the definition
\bel{eq:t}
t^\mu = t^{\mu\nu} k_\nu = \epsilon^{\mu \nu \alpha \beta} k_\nu u_\alpha \omega_\beta. 
\eel 
The four-vector $t$ is orthogonal to the vectors $u, k$ and $\omega$. In the local rest frame (LRF), where $u^\mu=(1,0,0,0)$, one finds that $\tv = \kv \times \omv$. Moreover, in \EQSM{eq:ext_Ts}{eq:GD_Ts} we have introduced the spin tensor $S^{\mu, \alpha \beta}_{\rm eq}$ defined by the expression
\bel{eq:spint_F}
S^{\mu, \alpha \beta}_{\rm eq} = u^\mu S^{\alpha \beta},
\eel
which is an analog of the perfect-fluid forms of $N^\mu_{\rm eq}$ and $T^{\lambda\mu}_{\rm eq}$ given by \EQSTWO{eq:NmuPF}{eq:TmunuPF}.

Equation~\EQn{eq:firstlaw_Ts} as a direct consequence implies the entropy conservation for a system that conserves baryon number, energy, linear momentum and spin, that is to say, the conservation laws $\p_\mu N^\mu_{\rm eq} =0$, $\p_\mu T^{\mu \lambda}_{\rm eq}=0$,  and $\p_\mu S^{\mu, \alpha\beta}_{\rm eq}~=~0$ imply $\p_\mu S^\mu_{\rm eq} =0$. We note that the spin conservation is a direct consequence of using a symmetric energy-momentum tensor in the considered formalism. In general, only the total angular momentum is conserved, 
\bel{eq:Jcon}
\p_\mu J^{\mu, \alpha\beta}=0,
\eel
with $J$ given as a sum of the orbital ($L$) and spin ($S$) parts
\bel{eq:totJ}
J^{\mu, \alpha\beta} = x^\alpha T^{\mu \beta} - x^\beta T^{\mu \alpha} + S^{\mu, \alpha\beta} \equiv L^{\mu, \alpha\beta} + S^{\mu, \alpha\beta}.
\eel
The last equation implies
\bel{eq:Snoncon}
\p_\mu S^{\mu, \alpha\beta}= T^{\beta \alpha}-T^{\alpha \beta}.
\eel
Thus, we find that the divergence of the spin tensor is determined by the antisymmetric part of the energy-momentum tensor. The latter vanishes in our case (for perfect fluid).

At this point it is important to note that the use of the expression \EQn{eq:spint_F} can be traced back to the very first model of a spinning fluid by Weyssenhoff and Raabe~\CITn{Weyssenhoff:1947iua}. It was also used in~\CIT{Florkowski:2017ruc}, where the first formulation of relativistic hydrodynamics for particles with spin $\onehalf$ was proposed. The form \EQn{eq:spint_F} has been subsequently used in many works that followed the methods of Israel and Stewart (positivity of the entropy current) to construct the framework of dissipative spin hydrodynamics~\CITn{Hattori:2019lfp, Fukushima:2020ucl, Sarwar:2022yzs, Daher:2022wzf, Xie:2023gbo, Daher:2024bah}. 

Despite the numerous applications of~\EQ{eq:spint_F}, its structure is not justified by microscopic calculations; see, for example, \CITS{DeGroot:1980dk,Florkowski:2017dyn,Florkowski:2018fap}. The way to solve this problem is to introduce generalized thermodynamic relations as in~\CIT{Florkowski:2024bfw}. In this work we further develop these ideas. 

%%%%%%%%%%%%%%%%%%%%%%%%%%%%%%%%%%%%%%%%%%%%%%
\section{Kinetic-theory results} 
\label{sec:kint}

In this section, we discuss the state of local thermodynamic equilibrium as defined within the framework of kinetic theory describing particles with spin~$\onehalf$. The framework of perfect spin hydrodynamics is based on a combination of two concepts: local equilibrium for particles with spin and the conservation laws for the baryon number, energy, linear momentum and the spin part of the angular momentum.

Our kinetic-theory approach uses a classical description of spin. It has been shown in~\CITn{Florkowski:2018fap} that for small values of the polarization tensor $\omega_{\mu\nu}$ the expression for the spin tensor obtained with classical treatment of spin is consistent with the result obtained from the calculations using a semiclassical expansion of the Wigner function. It was also found in~\CITn{Florkowski:2018fap}  that the classical spin treatment guarantees that the value of the Pauli-Lubański vector—as expected—never exceeds the value of $\sqrt{3/4}$. This is in contrast with the popular approach based on the Wigner function, which allows for infinite spin polarization~\footnote{This is due to the fact that the equilibrium spin part of the Wigner function is usually assumed to be of the form $\exp\left[-(i/2) \omega^{\mu\nu} \Sigma_{\mu\nu}\right]$, where $\Sigma_{\mu\nu} = (i/2) [\gamma_\mu,\gamma_\nu]$. That case was analyzed in detail in~\CIT{Florkowski:2017dyn}—see Eq.~(35) in~\CITn{Florkowski:2017dyn}, which leads to normalization problems; see also the discussion following Eq.~(50) in~\CIT{Florkowski:2018fap}.}. As a consequence, the classical treatment of spin is plausible to address the terms of order higher than one in the spin polarization tensor and is now one of the basic frameworks for construction of spin hydrodynamics. It has been subsequently used in numerous papers (see, for example, Refs.~\cite{Bhadury:2020puc, Bhadury:2020cop, Bhadury:2022ulr, Singh:2022ltu, Weickgenannt:2019dks, Weickgenannt:2021cuo, Weickgenannt:2020aaf, Wagner:2022amr, Weickgenannt:2022zxs, Weickgenannt:2023nge, Wagner:2024fhf, Hu:2021pwh}).

%%%%%%%%%%%%%%%%%%%%%%%%%%%%%%%%%%%%%%%%%%%%%%
\subsection{Classical spin description}

In the classical treatment of spin~\CITn{Mathisson:1937zz,2010GReGr..42.1011M}, one introduces the particle internal angular momentum tensor $s^{\alpha\beta}$ defined by the formula
\bel{eq:sab}
s^{\alpha\beta} = \f{1}{m} \epsUabgd p_\gamma s_\delta.
\eel
Here $p$ is the particle four-momentum satisfying the on-mass-shell condition $p^\mu p_\mu = m^2$ (with $m$ being the particle mass) and $s$ is the particle spin four-vector. Equation \EQn{eq:sab} implies that $s^{\alpha\beta} = -s^{\beta\alpha}$ and $s^{\alpha\beta} p_\beta = 0$. The spin four-vector is orthogonal to four-momentum $s \cdot p = 0$, hence we can write
\bel{eq:sa}
s^{\alpha} = \f{1}{2m} \,\epsUabgd p_\beta s_{\gamma \delta}.
\eel%
In the particle rest frame (PRF), where $p^\mu = (m,0,0,0)$, the four-vector $s^\alpha$ has only space components, $s^\alpha = (0,\sv_*)$, with the normalization $\spinl = \spin$. For particles with spin $\onehalf$ we use the value of the Casimir operator $\spin^2 = \onehalf \left( 1+ \onehalf  \right) = \threefourths$.

The basic object used in the kinetic theory is the phase-space distribution function $f(x,\pv)$. For particles with spin, $f(x,\pv)$ is generalized to a spin-dependent distribution $f(x,\pv,s)$. One commonly uses the notation $f(x,p,s)$ for $f(x,\pv,s)$ remembering that the energy $p^0$ is on the mass shell, namely, $p^0 = E_p =\sqrt{m^2 + \pv^2}$. Alongside the distribution function, we introduce the integration measures in momentum space
\bel{eq:dP}
\dd P = \f{\dd ^3p}{(2\pi)^3 E_p}
\eel
and spin space~\CITn{Florkowski:2018fap}
\bel{eq:dS}
\dd S = \f{m}{\pi \spin}  \, \dd ^4 s \, \delta(s \cdot s + \spin^2) \, \delta(p \cdot s).
\eel
More discussion of the measure and corresponding integrals is given in Appendix B.
%%%%%%%%%%%%%%%%%%%%%%%%%%%%%%%%%%%%%%%%%%%%%%%%%%%%%%%%%%%%%%%
\subsection{Equilibrium distribution functions and currents}

In local equilibrium, the spin-dependent distribution functions for particles ($+$) and antiparticles ($-$) have the Fermi-Dirac form
\bel{eq:fpm-FD}
f^{\pm_{\rm }}_{\rm eq}(x,p,s) = \LSB
\exp \LB
\mp \xi(x) + p \cdot \beta(x)  -  \frac{1}{2} \, \omega(x) : s \RB +1 \RSB^{-1}.
\eel 
We shall also use the compact notation
\bel{eq:fpm-FDy}
f^{\pm_{\rm }}_{\rm eq} = \f{1}{e^{y^\pm} + 1}
\eel 
with
\bel{eq:ypm}
y^\pm = \mp \xi(x) + p \cdot \beta(x) 
- \frac{1}{2} \omega(x) : s.
\eel
Equation \EQn{eq:fpm-FD} follows from the microscopic picture that baryon number, energy, linear momentum and the spin part of angular momentum are conserved in particle binary collisions. Similarly as the conservation of the baryon charge introduces a Lagrange multiplier that is the baryon chemical potential $\mu = T\xi$, the conservation of $s^{\alpha\beta}$ introduces six Lagrange multipliers grouped into a tensor spin chemical potential $\Omega_{\alpha\beta}= T \omega_{\alpha\beta}$. The arguments for this form of the local equilibrium function (in the case of Boltzmann statistics) were originally given in Sec.~6.3 of~\CIT{Florkowski:2018fap}. Subsequent important developments by the Frankfurt group, see Refs.~\CITn{Weickgenannt:2021cuo, Weickgenannt:2020aaf, Wagner:2022amr, Weickgenannt:2022zxs}, allowed for the inclusion of nonlocal collisions.

The macroscopic currents and tensors are obtained as moments of the distribution functions. In this way we obtain the baryon current
\bel{eq:Nmu}
N^\mu_{\rm eq} = \int \dd P \,\dd S \, p^\mu \, \left[f_{\rm eq}^+(x,p,s)-f_{\rm eq}^-(x,p,s) \right],
\eel
the energy-momentum tensor
\bel{eq:Tmunu}
T^{\mu \nu}_{\rm eq} = \int \dd P \,\dd S \, p^\mu p^\nu \, \left[f_{\rm eq}^+(x,p,s) + f_{\rm eq}^-(x,p,s) \right],
\eel 
and the spin tensor
\bel{eq:Slmunu}
\hspace{-0.5cm}S^{\lambda, \mu\nu}_{\rm eq}\!&=&\!\!\int \!\dd P \, \dd S \, \, p^\lambda \, s^{\mu \nu} 
\left[f_{\rm eq}^+(x,p,s)+ f_{\rm eq}^-(x,p,s) \right].
\eel
In addition, we define
\bel{eq:CNmu}
\mathcal{N}^\mu\!=\!-\int \dd P \,\dd S \, p^\mu \, \left[\ln(1-f_{\rm eq}^+)+\ln(1-f_{\rm eq}^-) \right].
\eel
In the traditional hydrodynamics, the current $\mathcal{N}^\mu$ can be directly expressed by local pressure and hydrodynamic flow, $\mathcal{N}^\mu = P \beta^\mu$, see Appendix~\ref{sec:pressFD}. However, this is not the case for spin hydrodynamics. Finally, we introduce the entropy current~\CITn{Landau:1980mil}
\begin{align}
\begin{split}\label{eq:Smu_FD}
S^{\mu}_{\rm eq} &= -\int \dd P \,\dd S \, p^\mu \, \left[f_{\rm eq}^+ \ln{f_{\rm eq}^+}-f_{\rm eq}^+\ln(1-f_{\rm eq}^+)+\ln(1-f_{\rm eq}^+)\right] \\&- \int \dd P \,\dd S \, p^\mu \left[ \, f_{\rm eq}^- \ln{f_{\rm eq}^-}-f_{\rm eq}^-\ln(1-f_{\rm eq}^-)+\ln(1-f_{\rm eq}^-) \right].
\end{split}
\end{align}
Using the identity
\bel{eq:nf1}
f_{\rm eq} \ln f_{\rm eq} - f_{\rm eq} \ln(1-f_{\rm eq})=  -y f_{\rm eq},
\eel
one can show that the entropy current $S^\mu_{\rm eq}$ can be expressed as a linear combination of other tensors and currents. Inserting \EQn{eq:nf1} into \EQn{eq:Smu_FD}, we obtain 
\begin{align}\begin{split}\label{eq:Smu_FD_Comp}
S^\mu_{\rm eq} 
&= \int \dd P \,\dd S \, p^\mu \LSB f_{\rm eq}^+ y^+ - f_{\rm eq}^- y^- \RSB 
-  \int \dd P \,\dd S \, p^\mu  \left[\ln(1-f_{\rm eq}^+)+\ln(1-f_{\rm eq}^-) \right] \\
&=  \int \dd P \,\dd S \, p^\mu \LSB f_{\rm eq}^+ \LB  
-\xi +p \cdot \beta - {\scriptstyle {\frac{1}{2}}} \omega: s \RB + f_{\rm eq}^- \LB  
\xi +p \cdot \beta - {\scriptstyle {\frac{1}{2}}} \omega : s \RB   \RSB + \mathcal{N}^\mu
\end{split}\end{align}
or
\bel{eq:Smu_FD1}
S^\mu_{\rm eq} = -N^\mu_{\rm eq} \xi + T^{\mu \alpha}_{\rm eq}\beta_\alpha - \frac{1}{2} S^{\mu,\alpha \beta}_{\rm eq}\omega_{\alpha \beta} + \mathcal{N}^\mu.
\eel
The last equation is almost identical to \EQ{eq:ext_Ts}, however, as noted above, the relation $\mathcal{N}^\mu = P \beta^\mu$ does not hold for particles with spin. This property has been recently emphasized in \CIT{Florkowski:2024bfw} for the case of the classical statistics. In this work, we find that \EQ{eq:Smu_FD1} holds also for the Fermi-Dirac statistics if the current $\mathcal{N}^\mu$ is defined by \EQ{eq:CNmu}. In the future it would be interesting to analyze if \EQ{eq:Smu_FD1} can be alternatively derived using the concepts introduced in \CIT{Jensen:2012jh}.

%%%%%%%%%%%%%%%%%%%%%%%%%%%%%%%%%%%%%%%%%%%%%%%%%%%%%%%%%%%%%%%%%%%%%%
\subsection{Generalized thermodynamic relations}
%%%%%%%%%%%%%%%%%%%%%%%%%%%%%%%%%%%%%%%%%%%%%%%%%%%%%%%%%%%%%%%%%%%%%%

For the distribution function of the form \EQn{eq:fpm-FDy}, one finds useful relations
\bel{eq:15}
\dd f_{\rm eq} \!=\! -\f{ e^y }{ (e^y+1)^2 }\dd y = -\f{1}{e^y+1} \, \f{e^y+1-1}{e^y+1} \, \dd y  = -f_{\rm eq}(1-f_{\rm eq})\dd y, 
\eel 
They can be used to obtain the expression for $d\mathcal{N}^\mu$ directly from \EQn{eq:CNmu}, namely,
\begin{align}\begin{split}\label{eq:CNmu_FD_CL}
\dd \mathcal{N}^{\mu}
&= -\dd \int \dd P \,\dd S \, p^\mu \, \left[\ln(1-f_{\rm eq}^+)+\ln(1-f_{\rm eq}^-) \right] =
\int \dd P \,\dd S \, p^\mu \, (f_{\rm eq}^+\!-\!f_{\rm eq}^-)\dd \xi \\
&- \!\!\!\int \dd P \,\dd S \, p^\mu p^\alpha (f_{\rm eq}^+\!+\!f_{\rm eq}^-) \dd \beta_{\alpha} + \frac{1}{2} \int \dd P \,\dd S \, p^\mu s^{\alpha\beta}(f_{\rm eq}^+\!+\!f_{\rm eq}^-)\dd \omega_{\alpha \beta},
\end{split}\end{align} 
which leads to
\bel{eq:dcalN}
\dd \mathcal{N}^{\mu} = N^{\mu}_{\rm eq}\dd \xi - T^{\mu \alpha}_{\rm eq}\dd \beta_{\alpha} + \frac{1}{2} S^{\mu,\alpha \beta}_{\rm eq}\dd \omega_{\alpha \beta}.
\eel
Finally, we use the form \EQn{eq:Smu_FD} to calculate $\dd S^\mu_{\rm eq}$. Combining the obtained result with $\dd \mathcal{N}^{\mu}$ given by \EQn{eq:dcalN}, we find
\bel{eq:dSmu}
\dd S^{\mu}_{\rm eq}= -\xi dN^\mu_{\rm eq} + \beta_{\alpha}\dd T^{\mu \alpha}_{\rm eq} -  \frac{1}{2} \omega_{\alpha \beta} \dd S^{\mu,\alpha \beta}_{\rm eq}.
\eel
Thermodynamic relations of the form \EQSTWOn{eq:dcalN}{eq:dSmu} have been recently derived for the case of classical statistics~\CITn{Florkowski:2024bfw}. Herein, we show that they are valid also for the Fermi-Dirac statistics, similarly as \EQ{eq:Smu_FD}. 

%%%%%%%%%%%%%%%%%%%%%%%%%%%%%%%%%%%%%%%%%%%%%%
\section{Small spin density expansion}
\label{sec:smallomega}

It should be emphasized that our derivation of Eqs.~\EQn{eq:dcalN} and \EQn{eq:dSmu} does not require that the spin polarization tensor be small. However, since the experimentally observed spin polarization effects typically are small, it makes sense to consider expansion in $\omega^{\mu\nu}$ or, alternatively, in $k^\mu$ and $\omega^\mu$ that are treated as small parameters (note that $\omega^{\mu\nu}$ is dimensionless in natural units). As pointed out in~\CIT{Florkowski:2024bfw}, a nontrivial and thermodynamically consistent treatment of spin can be achieved if we take into account at least quadratic corrections in $k$ and $\omega$. Thus, we expand the FD distribution functions around $y_s = 0$,
\bel{eq:expansion1}
\f{1}{e^{y^\pm_0+y_s}+1} &=& \f{1}{e^{y^\pm_0}+1}-\f{e^{y^\pm_0}}{(e^{y^\pm_0}+1)^2} \, y_s +\f{e^{y^\pm_0} (e^{y^\pm_0}-1)}{2 \, (e^{y^\pm_0}+1)^3} \, y_s^2 \, + \cdots . 
\eel
Here $y^\pm_0$ is the spin-independent part of $y^\pm$ defined by \EQ{eq:ypm}, 
\bel{eq:ypm0}
y^\pm_0 = \mp \xi(x) + p \cdot \beta(x),
\eel
while
\bel{eq:ys}
y_s = - \frac{1}{2} \, \omega(x) : s. 
\eel
With $f^\pm_0$ denoting the FD distribution with $\omega_{\alpha\beta}=0$, i.e., for the spinless case, we may write
\bel{eq:expansion2}
f_{\rm eq}^\pm = f^\pm_0 - f^\pm_0 (1-f^\pm_0) y_s +  \frac{1}{2}  f^\pm_0 (1-f^\pm_0) (1 - 2 f^\pm_0) y_s^2 + \cdots 
\eel
or 
\bel{eq:expansion3}
f_{\rm eq}^\pm = 
f^\pm_0 
+ \frac{1}{2} \, f^\pm_1  \, \omega : s 
+ \frac{1}{8} \,  f^\pm_2  \, (\omega : s)^2  + \cdots ,
\eel
where we have introduced the notation
\begin{align}\begin{split}\label{eq:f1}
 f^\pm_1 &= f^\pm_0 (1-f^\pm_0) = \f{e^{y^\pm_0}}{(e^{y^\pm_0}+1)^2} = \pm \f{\p}{\p \xi} f^\pm_0,\\
 f^\pm_2 &=f^\pm_0 (1-f^\pm_0) (1 - 2 f^\pm_0) = \f{e^{y^\pm_0} (e^{y^\pm_0}-1)}{(e^{y^\pm_0}+1)^3}
 = \f{\p^2}{\p \xi^2} f^\pm_0.
 \end{split}\end{align}

%%%%%%%%%%%%%%%%%%%%%%%%%%%%%%%%%%%%%%%%%%%%%%%%%%%%%%%%%%%%%%%%
\subsection{Baryon current}

The expansion in $\omega : s$  may be used in our definitions of the macroscopic currents. In this case, one can analytically integrate over the spin degrees of freedom first. Starting from the definition of the baryon current,
\bel{eq:NmuExp}
N_{\rm eq}^\mu = \int \dd P \,\dd S \, p^\mu \, \left[f_{\rm eq}^+(x,p,s)-f_{\rm eq}^-(x,p,s) \right],
\eel
and using the expansion \EQn{eq:expansion3} for the FD distribution functions as well as \EQ{eq:dSos}, we find
\bel{eq:01}
N_{\rm eq}^\mu = \int \dd P \,\dd S \, p^\mu 
\LB f^+_0 - f^-_0 \RB + \f{1}{8} \int \dd P \,\dd S \, p^\mu \LB f^+_2 - f^-_2 \RB (\omega:s)^2. 
\eel 
With the help of the spin integration rules \EQn{eq:dS2} and \EQn{eq:dSosos}, we further find
\begin{align}\begin{split}\label{eq:02}
N_{\rm eq}^\mu &=
2 \int \dd P \, p^\mu \LB f^+_0 - f^-_0 \RB 
+ \spin^2 \f{\omega:\omega}{6} \int \dd P \, p^\mu  
\LB f^+_2 - f^-_2 \RB \\ 
&+ \f{\spin^2 }{3m^2} \int \dd P  \, p^\mu p^\alpha p^\beta \omega^\gamma_{\phantom{\gamma}\alpha} \omega_{\beta \gamma} \LB f^+_2 - f^-_2 \RB .
\end{split}\end{align}
Here we introduce functions $Z^{\pm \alpha \beta ...}_{n}$, which appear in computations and are computed in Appendix A for several values of $n$,
\bel{}
Z^{\pm \alpha \beta ...}_{n} = \int \dd P p^{\alpha} p^{\beta}... f^{\pm}_n.
\eel
Then,  the baryon current can be written in a compact form as
\bel{eq:06}
N_{\rm eq}^\mu &=&  2(Z^{+\mu}_0- Z^{-\mu}_0) +  \spin^2\f{\omega:\omega}{6} (Z^{+\mu}_2- Z^{-\mu}_2) + \f{\spin^2}{3m^2} (   
Z^{+\mu\alpha\beta}_2 -   Z^{-\mu\alpha\beta}_2 )
\omega^\gamma_{\phantom{\gamma}\alpha} \omega_{\beta \gamma}.
\eel

%%%%%%%%%%%%%%%%%%%%%%%%%%%%%%%%%%%%%%%%%%%%%%%
\subsection{Energy-momentum tensor}

In the case of the energy-momentum tensor we use the expression
\begin{equation}\label{eq:TmunuExp}
T_{\rm eq}^{\mu \nu} = \int \dd P \,\dd S \, p^\mu p^\nu \, \left[f_{\rm eq}^+(x,p,s) + f_{\rm eq}^-(x,p,s) \right].
\end{equation}
Similarly, as in the case of the baryon current, the terms linear in the spin polarization tensor vanish. Thus, we obtain the following form of the energy-momentum tensor
\bel{eq:07}
T_{\rm eq}^{\mu\nu} &=& \int \dd P \,\dd S \, p^\mu p^\nu ( f_{0}^{+} + f_{0}^{+})  + \f{1}{8} \int \dd P \,\dd S \, p^\mu p^\nu (f^+_2+f^-_2 ) (\omega:s)^2.
\eel
Integration over the spin degrees of freedom yields
\begin{align}\begin{split}\label{eq:08}
T_{\rm eq}^{\mu\nu} &= 2 \int \dd P  \, p^\mu p^\nu (f^+_0+f^-_0) +   
\f{\spin^2}{6m^2} \int \dd P  \, p^\mu p^\nu  (f^+_2 + f^-_2)(m^2 \omega : \omega + 2 p^\alpha p^\beta \omega^\gamma_{\phantom{\gamma}\alpha} \omega_{\beta \gamma} )\\ 
&= 2 \int \dd P  \, p^\mu p^\nu (f^+_0+f^-_0) + 
\spin^2\f{\omega:\omega}{6} \int \dd P  \, p^\mu p^\nu  (f^+_2 + f^-_2) \\
&+ \f{\spin^2}{3m^2}\int \dd P p^\mu p^\nu p^\alpha p^\beta (f^+_2 + f^-_2) \omega^\gamma_{\phantom{\gamma}\alpha} \omega_{\beta \gamma},
\end{split}\end{align}
which can be rewritten in terms of the tensors $Z$, namely
as
\bel{eq:09}
T_{\rm eq}^{\mu \nu} = 
2 \LB Z^{+\mu \nu}_{0}\!+\!Z^{-\mu \nu}_{0} \RB
\!+\! \spin^2 \,\f{\omega:\omega}{6} 
\LB Z^{+\mu \nu}_{2}\!+\!Z_{2}^{-\mu \nu} \RB 
\!+\! \f{\spin^2}{3m^2} \,
\LB Z^{+\mu \nu \alpha \beta}_{2}\!+\!Z^{-\mu \nu \alpha \beta}_{2} \RB\omega^\gamma_{\phantom{\gamma}\alpha} \omega_{\beta \gamma}.
\eel
%

%%%%%%%%%%%%%%%%%%%%%%%%%%%%%%%%%%%%%%%%%%%%%%%
\subsection{Spin tensor}

In the case of the spin tensor, we use the formula
\bel{eq:SlmunuExp}
S_{\rm eq}^{\lambda, \mu\nu} &=& \int \!\dd P \, \dd S \, \, p^\lambda \, s^{\mu \nu} 
\left[f_{\rm eq}^+(x,p,s)+ f_{\rm eq}^-(x,p,s) \right] 
\eel
and expand the FD distribution functions up to linear terms in the spin polarization tensor, which leads to the expression
\begin{align}
\begin{split}\label{eq:11}
S_{\rm eq}^{\lambda, \mu\nu} \!&=\!\! \f{1}{2} \int \!\dd P \, \dd S \, \, p^\lambda \, s^{\mu \nu} ( f^+_1 + f^-_1 )(\omega : s) \\ 
&=
\f{2 \spin^2}{3} \int \dd P p^\lambda (f_{1}^+ + f_1^-)\omega^{\mu \nu} + 
\frac{2 \spin^2}{3 m^2} \int \dd P p^\lambda p^\alpha (p^\mu \omega^{\nu}_{\,\,\, \alpha}
-p^\nu \omega^{\mu}_{\,\,\, \alpha})(f^+_1 + f^-_1).
\end{split}
\end{align}
Then, one obtains the compact expression
\begin{align}
\begin{split}\label{eq:13}
S_{\rm eq}^{\lambda,\mu \nu} = & \f{2 \spin^2 }{3} \,  \omega^{\mu \nu} \,(Z^{+\lambda}_1 +   Z^{-\lambda}_1) + \f{2 \spin^2 }{3m^2} \LSB  \, ( Z^{+ \lambda \alpha \mu }_{1} + Z^{- \lambda \alpha \mu }_{1} )\omega^{\nu}_{\,\,\, \alpha}  -  ( Z^{+ \lambda \alpha \nu }_{1} + Z^{- \lambda \alpha \nu }_{1} )\omega^{\mu}_{\,\,\, \alpha} \RSB.
\end{split}
\end{align}
We note that the baryon current and the energy-momentum tensor contain even terms in the expansion, while the spin tensor includes the odd terms. However, because in the thermodynamic relations the spin tensor is multiplied by the spin polarization tensor, only even terms in $\omega_{\alpha \beta}$ appear there. 
%%%%%%%%%%%%%%%%%%%%%%%%%%%%%%%%%%%%%%%%%%%%%%%%
% BOLTZMANN
%%%%%%%%%%%%%%%%%%%%%%%%%%%%%%%%%%%%%%%%%%%%%%%%
\section{Boltzmann Statistics}
\label{sec:Boltzmann}

Since the  explicit expressions for the tensors $Z^{\pm \alpha\beta \ldots}_{n}$ are rather complicated, in the next sections we restrict our considerations to the Boltzmann statistics. We come back to the discussion of the  FD case in Appendix~\ref{sec:Ztensors}. The classical statistics is obtained by neglecting the term $+1$ in~\EQ{eq:fpm-FD}. In this way we obtain the Boltzmann distribution
\bel{eq:fpm-spin-B}
f_{\rm eq}^{\pm_{\rm }}(x,p,s) = \exp \LSB \pm\xi(x)-p \cdot \beta(x)
+ \frac{1}{2} \omega(x) : s \RSB. 
\eel 
In addition, as the classical limit corresponds to a dilute system, we can always neglect $f$ in expressions such as $(1-f)$ or $(1-2f)$. Thus, \EQSM{eq:expansion2}{eq:expansion3} become
\bel{eq:expansion2B}
f_{\rm eq}^\pm = f^\pm_0 - f^\pm_0 y_s +  \frac{1}{2}  f^\pm_0  y_s^2 + \cdots 
\eel
or
\bel{eq:expansion3B}
f_{\rm eq}^\pm = 
f^\pm_0 \LSB 1  + \frac{1}{2} \, \omega : s  
+ \frac{1}{8} \,  (\omega : s)^2 + \cdots \RSB.
\eel
Clearly, the terms in the squared brackets directly come from the expansion of the exponential function \EQn{eq:fpm-spin-B}.

%%%%%%%%%%%%%%%%%%%%%%%%%%%%%%%%%%%%%%%%%%%%%%%%%%%%%%%%%%%
\subsection{Baryon and particle currents}

Factorization of the Boltzmann distribution function into the ``momentum'' and ``spin'' parts makes the calculation of macroscopic quantities quite straightforward. First we integrate over the spin degrees of freedom and then over momentum, which leads to expressions involving modified Bessel functions. In the case of the baryon current, including terms up to the second order, we find
\begin{align}\begin{split}\label{eq:Computation_Nmu}
N_{\rm eq}^{\mu} 
&= 2\sinh{\xi}\int \dd P\, p^{\mu}e^{-p\cdot \beta}  \LSB  2 + \f{1}{8} \int dS (\omega : s )^2 \RSB \\
&= 4\sinh{\xi}\int \dd P\, p^{\mu}e^{-p\cdot \beta} \LSB \LB 1 +\frac{\spin^{2}}{12} \omega:\omega \RB +\frac{\spin^{2}}{6m^{2}}p^{\alpha}p^{\beta}\omega^{\gamma}_{\phantom{\gamma} \alpha}\omega_{\beta \gamma} \RSB \\  
&= 4\sinh{\xi} \LB 1+\frac{ \spin^{2}}{12} \, \omega:\omega \RB \underset{Z^{\mu}}{\underbrace{\int \dd P\, p^{\mu}e^{-\beta \cdot p}}}  +
 \frac{2 \textgoth{s:}^{2}\sinh{\xi}}{3 m^{2}}\underset{Z^{\mu \alpha \beta}}{\underbrace{\int \dd P\, p^{\mu} p^{\alpha} p^{\beta}e^{-\beta \cdot p}}} \omega^\gamma_{\phantom{\gamma}\alpha} \omega_{\beta \gamma}. 
\end{split}\end{align}
We note that this formula is the classical (Boltzmann) limit of~\EQ{eq:06}. The integration over the spin degrees of freedom is done according to the rules \EQSTWOn{eq:dSos}{eq:dSosos}, see~\CIT{Florkowski:2018fap}. In the last line of~\EQ{eq:Computation_Nmu}, we underlined the integrals that define the tensors $Z^\mu$ and $Z^{\mu \alpha \beta}$. Clearly, this type of tensors represents classical limit of the tensors $Z^{\pm \alpha\beta \ldots}_{n}$ defined above in the context of the FD statistics, 
\bel{eq:ZFDZB}
Z^{\pm \alpha\beta \ldots}_{n} \quad \longrightarrow \quad e^{\pm \xi} Z^{\alpha\beta \ldots}, \qquad n=0,1,2.
\eel 
The explicit forms of $Z^\mu$ and $Z^{\mu \alpha \beta}$ can be found in \CIT{Cercignani:2002rh}~\footnote{Note that the expressions given in \CIT{Cercignani:2002rh} should be divided by $(2\pi)^3$, since our integration measure $dP$ includes this extra factor in the denominator; compare~\EQ{eq:dP}.}. In the special case of $Z^\mu$ we have 
\bel{eq:Zmu}
Z^\mu = \f{T^3}{2\pi^2} z^2  K_2(z) u^\mu.
\eel
Here and below the functions $K_n(z)'s$ are the modified Bessel functions of the second kind with the argument $z~=~m/T$. After performing all the necessary tensor contractions, (see Eqs.~\EQn{eq:RN1}--\EQn{eq:Z3oo} in Appendix B for details), this leads to the decomposition
\bel{eq:NmuG}
N^\mu = {\bar n} u^\mu + n_t t^\mu = ( n_{0}+ n_{2}^{k} +  n_{2}^{\omega} )u^{\mu}+n_{t}t^{\mu},
\eel
where the coefficients $n_{0},  n_{2}^{k},  n_{2}^{\omega}$, and $n_t$ have the following forms:
\begin{align}\begin{split}
n_{0}&=\f{2\sinh{\xi}}{\pi^{2}}z^{2}T^{3}K_{2}(z), \quad
{n}_{2}^{k} =-\f{2 \spin^{2} \sinh{\xi}}{3\pi^{2}}   z T^{3}K_{3}(z)k^{2},\\
{n}_{2}^{\omega} &= -\f{\spin^{2}\sinh{\xi}}{3\pi^{2}} z T^{3} \LSB z K_{2}(z) + 2 K_{3}(z) \RSB \omega^{2}, \quad  n_{t}=-\frac{2\spin^{2}\sinh{\xi}}{3\pi^2}  z T^{3} K_{3}(z).
\end{split}\end{align}
The coefficient $n_0$ describes the baryon density of a relativistic spinless gas.

In the case of classical statistics, the current $\mathcal{N}^\mu$ becomes the sum of particle and antiparticle currents, namely
\bel{eq:BNmu}
\mathcal{N}^\mu\!=\!\int \dd P \,\dd S \, p^\mu \, \left[f_{\rm eq}^+(x,p,s)+f_{\rm eq}^-(x,p,s) \right].
\eel
Hence, a simple relation holds
\bel{eq:calN}
{\cal N}^\mu = \coth\xi\,\,N_{\rm eq}^\mu .
\eel
We note that throughout this work we assume that $\mu \neq 0$ ($\xi \neq 0)$.

%%%%%%%%%%%%%%%%%%%%%%%%%%%%%%%%%%%%%%%%%%%%%%%%%%%%%%%%%%%%%%%%%%%%%%%%%%%%
% ENERGY-MOMENTUM TENSOR
%%%%%%%%%%%%%%%%%%%%%%%%%%%%%%%%%%%%%%%%%%%%%%%%%%%%%%%%%%%%%%%%%%%%%%%%%%%%
\subsection{Energy-momentum tensor}

Using the definition of the energy-momentum tensor \EQn{eq:Tmunu} and expanding the spin part of the distribution functions up to the second order in the spin polarization tensor, we obtain the formula
\begin{align}\begin{split}\label{eq:TmunuZ}
T_{\rm eq}^{\mu \nu} &= 4\cosh{\xi} \left( 1 +\frac{\spin^{2}}{12} \omega:\omega \right)\underset{Z^{\mu \nu}}{\underbrace{\int \dd P\, p^{\mu} p^{\nu}e^{-\beta \cdot p}}} \\
&+ \frac{2 \spin^{2} \cosh{\xi} }{3 m^{2}} \underset{Z^{\mu \nu \alpha \beta}}{\underbrace{\int \dd P\, p^{\mu} p^{\nu} p^{\alpha} p^{\beta}e^{-\beta \cdot p}}} 
\, \omega^{\gamma}_{\phantom{\gamma}\alpha} \omega_{\beta \gamma}.
\end{split}\end{align}
Here we underlined the tensors $Z^{\mu\nu}$ and $Z^{\mu\nu\alpha\beta}$, whose explicit forms are given in~\CITn{Cercignani:2002rh}.
Combining all the expressions \EQn{eq:RT1} with \EQn{eq:Z4oo}, we find the formula
\begin{align}\begin{split}\label{eq:Tmunu_decomposed}
T_{\rm eq}^{\mu \nu} &=
{\bar \varepsilon} u^\mu u^\nu - {\bar P} \Delta^{\mu\nu} + P_k \, k^\mu k^\nu   + P_\omega \,\omega^\mu \omega^\nu + P_t \,(t^\mu u^\nu + t^\nu u^\mu) \\
&= (\varepsilon_{0}+{\varepsilon}_{2}^{k}+{\varepsilon}_{2}^{\omega})u^{\mu}u^{\nu} - ({P}_{0}+{P}^{k}_{2}+{P}^{\omega}_{2})\Delta^{\mu\nu}+ \\
&+ P_{k}k^{\mu}k^{\nu}+P_{\omega}\omega^{\mu}\omega^{\nu}+P_{t}(t^{\mu}u^{\nu}+t^{\nu}u^{\mu}),
\end{split}\end{align}
where the coefficient functions read
\begin{align}\begin{split}\label{coef:epsilon_0}
{\varepsilon}_{0} &=\frac{2\cosh{\xi}}{\pi^{2}}z^{2}T^{4} \LSB z K_{3}(z)-K_{2}(z) \RSB,\\
{\varepsilon}_{2}^{k} &= -\frac{2\spin^{2}\cosh{(\xi)}}{3 \pi^{2}}z T^{4}\LSB z K_{2}(z) + 5K_{3}(z) \RSB k^{2}, \\
{\varepsilon}_{2}^{\omega} &= -\f{\spin^{2}\cosh{(\xi)}}{3\pi^{2}} zT^{4}\LSB z K_{2}(z)+(z^{2}+10)K_{3}(z) \RSB\omega^{2}
\end{split}\end{align}
and
\begin{align}\begin{split}\label{eq:pressure}
{P_{0}}&=\frac{2\cosh{\xi}}{\pi^{2}}z^{2}T^{4} K_{2}(z),\\
{P}^{k}_{2}&=-\frac{4 \spin^{2}\cosh{\xi}}{3 \pi^{2}}zT^{4}K_{3}(z)k^{2}, \quad {P}^{\omega}_{2} = -\frac{\spin^{2}\cosh{\xi}}{3\pi^{2}} zT^{4}\LSB z K_{2}(z)+4K_{3}(z) \RSB\omega^{2},\\
P_{t}&=\f{2 \spin^{2} \cosh{\xi}}{3\pi^{2}}zT^{4} \LSB K_{3}(z)-zK_{4}(z) \RSB, \quad  P_{k}=P_{\omega}=-\f{2 \spin^{2}\cosh{\xi}}{3\pi^{2}} z \, T^{4} K_{3}(z).
\end{split}\end{align}
Obviously, the quantities $\varepsilon_0$ and $P_0$ correspond to the energy density and pressure of spinless particles, respectively. We also have $P_0 = \coth\xi \, n_0 T$, which is the relativistic version of the Clapeyron equation. We note that the energy-momentum tensor can be rewritten as
\begin{align}\begin{split}\label{eq:TmunuGt}
T^{\mu\nu}_{\rm eq} &= {\bar \varepsilon}(T,\xi, k^2, \omega^2) u^\mu u^\nu - {\bar P}_{k\omega}(T,\xi, k^2, \omega^2) \Delta^{\mu\nu} \\
&+ P_{k\omega}(T,\xi) (\, k^{\LAB \mu} k^{\nu \RAB}  +\,\omega^{\LAB \mu} \omega^{\nu \RAB} ) + P_t(T,\xi) \,(t^\mu u^\nu + t^\nu u^\mu),
\end{split}\end{align}
where ${\bar P}_{k\omega} = {\bar P} - (1/3)  P_{k\omega} (k^2 + \omega^2)$. 

%%%%%%%%%%%%%%%%%%%%%%%%%%%%%%%%%%%%%%%%%%%%%%%%%%%%%%%%%%%%%%%%%%%%%%%%%%%%
% SPIN TENSOR
%%%%%%%%%%%%%%%%%%%%%%%%%%%%%%%%%%%%%%%%%%%%%%%%%%%%%%%%%%%%%%%%%%%%%%%%%%%%
\subsection{Spin tensor}

In the next step, we consider the spin tensor $S^{\lambda,\mu \nu}_{\rm eq}$ given by \EQ{eq:Slmunu}. Using \EQ{eq:dSsos}, we can again express it in terms of the tensors $Z$,
\bel{eq:Slmunu1}
S_{\rm eq}^{\lambda,\mu \nu} &=& 2\cosh{\xi} \int \dd P\, p^{\lambda}e^{-p \cdot \beta}
\int \dd S s^{\mu \nu}\LSB 1+\f{1}{2} \omega : s \RSB \nn \\
&=& \cosh{\xi}\int \dd P\, p^{\lambda} e^{-p \cdot \beta} \, \int \dd S \, s^{\mu \nu} \,  \omega : s\\
&=& \f{4\spin^{2} \cosh{\xi} }{3m^{2}} \LSB m^{2}\omega^{\mu \nu}\underset{Z^{\lambda}}{\underbrace{\int \dd P\, p^{\lambda} e^{-\beta \cdot p}}} +
\omega^{\nu}_{\phantom{\nu}\alpha}\underset{Z^{\lambda \alpha \mu}}{\underbrace{\int \dd P\, p^{\lambda}p^{\alpha}p^{\mu} e^{-\beta \cdot p}}} - \omega^{\mu}_{\phantom{\mu}\alpha}\underset{Z^{\lambda \alpha \nu}}{\underbrace{\int \dd P\, p^{\lambda}p^{\alpha}p^{\nu} e^{-\beta \cdot p}}} \RSB. \nn
\eel
To obtain the final form, we need an explicit expression for the contraction \EQn{eq:ZZ}. We also introduce the tensor
\begin{align}\begin{split}\label{eq:tlmunu}
t^{\lambda\mu\nu} &= \omega^{\nu\lambda} u^{\mu} - \omega^{\mu\lambda} u^{\nu}
+ g^{\lambda\mu} k^{\nu} - g^{\lambda\nu} k^{\mu} \\
&= u^\lambda \LB u^\mu k^\nu - u^\nu k^\mu \RB + u^\mu t^{\nu\lambda} - u^\nu t^{\mu\lambda} +  g^{\lambda\mu} k^{\nu} - g^{\lambda\nu} k^{\mu}.
\end{split}\end{align}
Then, the spin tensor $S_{\rm eq}^{\lambda,\mu \nu}$ can be written as
\bel{eq:Slmunu_final}
S_{\rm eq}^{\lambda, \mu \nu}=A_{1}u^{\lambda}\omega^{\mu \nu}
+\f{A_{2}}{2} u^{\lambda} \LB u^{\mu}k^{\nu} - u^{\nu}k^{\mu} \RB
+\f{A_{3}}{2} \, t^{\lambda \mu \nu},
\eel
where
\begin{align}\begin{split}\label{eq:A1_A2_A3}
A_{1} &= \f{2\spin^{2}\cosh{\xi}}{3\pi^{2}}zT^{3} \LSB zK_{2}(z)+2K_{3}(z)\RSB, \\
A_{2} &= \f{4\spin^{2}\cosh{\xi}}{3\pi^{2}}z^{2}T^{3}K_{4}(z), \\
A_{3} &= -\f{4\spin^{2}\cosh{\xi}}{3\pi^{2}}z T^{3}K_{3}(z) .
\end{split}\end{align}
This result is consistent with the decomposition used in \CIT{Florkowski:2019qdp}. It is also convenient to introduce a coefficient $A$ defined by the expression
\bel{eq:A}
A = A_1 - \f{A_2}{2}-A_3 = -\f{4\spin^{2}\cosh{\xi}}{3\pi^{2}}z T^{3}K_{3}(z) = A_3, 
\eel
which allows us to rewrite the spin tensor in a compact form as
\bel{eq:spint_GLW}
S_{\rm eq}^{\lambda, \mu \nu} = u^\lambda 
\LSB A \LB k^\mu u^\nu - k^\nu u^\mu \RB + A_1 t^{\mu\nu}
\RSB  + \frac{A}{2} 
\LB t^{\lambda \mu} u^\nu - t^{\lambda \nu} u^\mu + 
\Delta^{\lambda \mu} k^\nu - \Delta^{\lambda \nu} k^\mu 
\RB.
\eel
An important feature of \EQ{eq:spint_GLW} is that it does not agree with the phenomenological version~\EQn{eq:spint_F}. Moreover, even if the term on the right-hand side of \EQ{eq:spint_GLW} that is orthogonal to $u^\lambda$ is neglected, still the spin density tensor defined by \EQ{eq:spint_GLW} cannot be written in the form $S^{\mu\nu}= A \, \omega^{\mu\nu}$, as the coefficients $A$ and $A_1$ are different. 

As the matter of fact, the spin equation of state of the form $S^{\mu\nu}= A \, \omega^{\mu\nu}$ has been recently analyzed and excluded \CITn{Daher:2024ixz} as leading to unstable behavior of rest frame modes in the first-order \CITn{Sarwar:2022yzs,Daher:2022wzf} and second-order \CITn{Xie:2023gbo,Daher:2024bah} dissipative spin hydrodynamics. The conclusion of these works is that the spin density tensor $S^{\mu\nu}$ should depend differently on the electriclike and magneticlike components of the spin tensor, with the electric susceptibility being negative and the magnetic being positive. We emphasize that this is really the case for our kinetic expression \EQn{eq:spint_GLW} since $A <0$ and $A_1 > 0$. In particular, for small values of $z = m/T$ we find
\bel{eq:A1overA}
\f{A_1}{A} = -1 - \f{z^2}{8} + \cdots ,
\eel
while for large $z = m/T$ we have
\bel{eq:AoverA1}
\f{A}{A_1} = -\f{2}{z} - \f{1}{z^2} + \cdots .
\eel
Thus, for small $z$ the coefficients $A$ and $A_1$ have almost the same magnitude but opposite signs, whereas for large $z$ the coefficient $A$ is much smaller than $A_1$. In the latter case, the magneticlike component of the spin density tensor dominates the system's behavior, as first noted in~\CIT{Florkowski:2017dyn}. 

%%%%%%%%%%%%%%%%%%%%%%%%%%%%%%%%%%%%%%%%%%%%%%%%%%%%%%%%%%%%%%%%%%
\subsection{Entropy current}
In addition to $N^\mu$, $\mathcal{N}^{\mu}$, $T^{\mu \nu}$, and $S^{\lambda, \mu\nu}$, we introduce the entropy current using the standard Boltzmann definition~\CITn{Landau:1980mil}
\bel{eq:Hmu1}
S_{\rm eq}^\mu =\!-\!\int \dd P \, \dd S \, p^\mu 
\LSB 
f_{\rm eq}^+ \LB \ln f_{\rm eq}^+\!\!-\!1\RB\!+\! 
f_{\rm eq}^- \LB \ln f_{\rm eq}^-\!\!-1\!\RB \RSB,
\eel
which directly leads to the formula~\CITn{Florkowski:2019qdp}
\bel{eq:Hmu2}
S_{\rm eq}^\mu =  T_{\rm eq}^{\mu \alpha} \beta_\alpha-\f{1}{2} \omega_{\alpha\beta} S_{\rm eq}^{\mu, \alpha \beta}
-\xi N_{\rm eq}^\mu + {\cal N}^\mu.
\eel
This form provides us with the general form of the entropy current, after subsequent computations of entropy current parts, namely
\bel{eq:Tmunua}
T_{\rm eq}^{\mu \alpha} u_{\alpha} = \bar{\varepsilon} u^{\mu} + P_{t} t^{\mu}
\eel
and
\bel{eq:14}
-\xi N_{\rm eq}^{\mu}+\mathcal{N}^{\mu} = 
\LB \coth \xi - \xi \RB N_{\rm eq}^{\mu} =  
\LB \coth \xi - \xi \RB \bar{n} u^{\mu} 
+ \LB \coth \xi - \xi \RB n_{t}t^{\mu},
\eel
together with
\bel{eq:Somega}
\f{1}{2} \omega_{\alpha\beta}  S_{\rm eq}^{\mu, \alpha \beta} = u^\mu (A k^2 - A_1 \omega^2) + A_3 t^\mu 
\equiv {\bar s} u^\mu + s_t t^\mu.
\eel 
Then, we obtain the following form of the entropy current 
\bel{eq:NSmuG1}
S_{\rm eq}^\mu = {\bar \sigma} u^\mu + \sigma_t t^\mu,
\eel
where the coefficients are
\bel{eq:sigmas}
\bar{\sigma}=\frac{\bar{\varepsilon}}{T}+(\coth \xi - \xi)\bar{n}-\bar{s}, \hspace{0.7cm} \sigma_{t}=\f{P_{t}}{T} +  (\coth{\xi}-\xi )n_{t}-s_{t}.
\eel

%%%%%%%%%%%%%%%%%%%%%%%%%%%%%%%%%%%%%%%%%%%%%%
\section{Close-to-equilibrium behavior}
\label{sec:closetoeq}

%%%%%%%%%%%%%%%%%%%%%%%%%%%%%%%%%%%%%%%%%%%%%%
\subsection{Nonequilibrium entropy current}  

To extend the formalism presented above to a theory including dissipative phenomena, we follow the method initiated by Israel and Stewart~\CITn{Israel:1979wp}. It relies on the replacements of the equilibrium currents $N^\mu_{\rm eq}$, $T^{\mu \alpha}_{\rm eq}$ and $S^{\mu, \alpha \beta}_{\rm eq}$ in \EQ{eq:Hmu2} by the general nonequilibrium expressions that can be represented as the equilibrium terms plus nonequilibrium corrections: $N^\mu = N^\mu_{\rm eq} + \delta N^\mu$, $T^{\mu \alpha} = T^{\mu \alpha}_{\rm eq} + \delta T^{\mu \alpha}$ and $S^{\mu, \alpha \beta} = S^{\mu, \alpha \beta}_{\rm eq} +\delta S^{\mu, \alpha \beta}$. In this way we arrive at the formula
\bel{eq:HmuN}
S^\mu =  T^{\mu \alpha} \beta_\alpha-\f{1}{2} \omega_{\alpha\beta} S^{\mu, \alpha \beta}
-\xi N^\mu + {\cal N}^\mu.
\eel
Next, we calculate the divergence of the entropy current defined by \EQ{eq:HmuN}. Since in the general (nonequilibrium) case the energy-momentum tensor contains nonsymmetric parts,  we should use the equations~\footnote{Only the total angular momentum is conserved in the general case ($\p_\mu J^{\mu, \alpha\beta}=0$ with $J^{\mu, \alpha\beta} = x^\alpha T^{\mu \beta} - x^\beta T^{\mu \alpha} + S^{\mu, \alpha\beta}$), which is why we have $\p_\mu S^{\mu, \alpha\beta}= T^{\beta \alpha}-T^{\alpha \beta}$. }
\bel{eq:con_eqN}
\p_\mu N^\mu=0, \qquad 
\p_\mu T^{\mu \nu}=0, \qquad
\p_\mu S^{\mu, \alpha \beta} 
= T^{\beta \alpha} - T^{\alpha \beta} .
\eel
This leads to the following expression for the entropy production
\bel{eq:divS}
\p_\mu S^\mu =  
- \delta N^\mu \p_\mu \xi
+ \delta T^{\mu \lambda}_s \p_\mu \beta_\lambda 
+ \delta T^{\mu \lambda}_a \LB \p_\mu \beta_\lambda 
- \omega_{\lambda\mu} \RB
-\f{1}{2} \delta S^{\mu, \alpha \beta} \p_\mu \omega_{\alpha\beta} ,
\eel
where the labels $s$ and $a$ denote the symmetric and antisymmetric parts of the energy-momentum tensor. 

Equation \EQn{eq:divS} implies that the {\it global equilibrium} is defined by the generalized Tolman-Klein conditions~\CITn{Tolman:1934,Klein:1949}, which include the two standard equations, $\p_\mu \xi = 0$ and $\p_{( \mu} \beta_{\lambda )} = 0$, and an extra constraint that the spin polarization tensor is given by the thermal vorticity, $\omega_{\lambda\mu} = \p_{[ \mu} \beta_{\lambda ]}$. Nevertheless, in {\it local equilibrium} $\omega_{\lambda\mu}$ and $\p_{[ \mu} \beta_{\lambda ]}$ are not directly related and may be significantly different from each other. In this respect, we differ from the concept of local equilibrium originally proposed in~\CITn{Becattini:2013fla}. In our approach, the behavior of the spin polarization tensor, $\omega_{\mu\nu} = \Omega_{\mu\nu}/T$, is similar to the behavior of the ratio $\xi = \mu/T$ in standard (spinless) relativistic hydrodynamics. In global equilibrium $\xi =$~const., while in local equilibrium a direct connection between $T$ and $\mu$ is lost. 

Although \EQ{eq:divS} or its special case for $\xi=0$ was obtained before (see, for example: Eq.~(10) in \CITn{Hattori:2019lfp}, (23) in \CITn{Fukushima:2020ucl}, (21) in \CITn{Biswas:2023qsw}, and the QFT analysis in \CITn{Becattini:2023ouz}), the previous studies considered always the local equilibrium state without the orthogonal corrections discussed above. Thus, it is important to extend the previous analyses by considering a different reference point for local equilibrium quantities.

%%%%%%%%%%%%%%%%%%%%%%%%%%%%%%%%%%%%%%%%%%%%%%
\subsection{General tensor decompositions} 

To establish the form of the deviations $\delta N^\mu$, $\delta T^{\mu \alpha}$, and $\delta S^{\mu, \alpha \beta}$, we need the equilibrium terms defined above and the most general forms of the tensors $N^\mu$, $T^{\mu \alpha} $, and $S^{\mu, \alpha \beta}$. The latter are usually obtained by making sequential projections along $u^\mu$ and separation of the symmetric and antisymmetric tensors. 

Writing the baryon nonequilibrium current as $N^\mu = N^\alpha g^{\mu}_{\,\,\,\alpha} =  N^\alpha (\Delta^{\mu}_{\,\,\,\alpha} + u^\mu u_\alpha)$, where $u$ is an arbitrary time-like vector, we obtain a decomposition
\bel{eq:Ndec}
N^\mu = a u^\mu + b^\mu,
\eel
where $b^\mu u_\mu = 0$. Through more algebra, we obtain an analogous expression for the energy-momentum tensor~\CITn{Hattori:2019lfp}
\begin{align}\begin{split}\label{eq:Tdec}
T^{\mu\nu} &= c u^\mu u^\nu + d^\mu_s u^\nu + d^\nu_s u^\mu  + d^\mu_a u^\nu - d^\nu_a u^\mu + e^{\mu\nu}_a + e^{\mu\nu}_s,
\end{split}\end{align}
where $d^\mu_s u_\mu = d^\mu_a u_\mu = e^{\mu\nu}_a u_\mu = e^{\mu\nu}_s u_\mu = 0$, and $e^{\mu\nu}_a$ ($e^{\mu\nu}_s$) is an antisymmetric (symmetric) part of $e^{\mu\nu}$. Extracting the traceless part of $e^{\mu\nu}_s$, we can rewrite \EQ{eq:Tdec} as
\begin{align}\begin{split}\label{eq:TdecT}
T^{\mu\nu} &= c u^\mu u^\nu - e \Delta^{\mu\nu}
+ d^\mu_s u^\nu + d^\nu_s u^\mu + e^{\LAB \mu\nu \RAB}_s 
+ d^\mu_a u^\nu - d^\nu_a u^\mu + e^{\mu\nu}_a,
\end{split}\end{align}
where $e = -(1/3) e^\lambda_{s\,\,\lambda}$. The parametrization \EQn{eq:TdecT} involves 19 independent parameters: the scalars $c$ and $e$ introduce 2 parameters, the vectors $u^\mu$, $d_s^\mu$, $d_a^\mu$ bring 9 parameters (due to the normalization and orthogonality conditions), the tensors $e^{\LAB \mu\nu \RAB}_s$ and $e^{\mu\nu}_a$ have 5 and 3 independent components, respectively. Since $T^{\mu\nu}$ may have 16 independent components, we can eliminate 3 degrees of freedom by the so-called frame choice for the hydrodynamic flow~$u^\mu$.\footnote{We do not use the freedom of choosing a specific form of the hydrodynamic flow $u^\mu$ here, since it is not important for our conclusions.}

Once again with the same simple algebraic method, we obtain a decomposition of the spin tensor~\CITn{Becattini:2011ev,Biswas:2023qsw}
\begin{align}\begin{split}\label{eq:Sdec}
S^{\lambda, \mu\nu} &= u^\lambda \LSB \LB f^\mu u^\nu - f^\nu u^\mu \RB
+ h^{\mu \nu} \RSB + i^{\lambda \mu} u^\nu - i^{\lambda \nu} u^\mu
+ j^{\lambda\mu\nu}.
\end{split}\end{align}
Note that the decomposition of the spin tensor is determined by the fact that $S^{\lambda, \mu\nu}$ must be antisymmetric in the last two indices. Thus, the vector and tensor fields introduced above fulfill the following constraints: $f^\mu u_\mu = 0$, $h^{\mu \nu} = - h^{\nu \mu}$ with $h^{\mu \nu} u_\mu = 0$, $i^{\lambda \mu} u_\lambda = i^{\lambda \mu} u_\mu = 0$, and $j^{\lambda\mu\nu} = - j^{\lambda\nu\mu} $ with $j^{\lambda\mu\nu} u_\lambda = j^{\lambda\mu\nu} u_\mu = j^{\lambda\mu\nu} u_\nu = 0$. In general, the spin tensor has 24 parameters: $f^\mu$ and $h^{\mu \nu}$ bring 3 independent components each, $i^{\lambda \mu}$ has 9 independent components (3 in the antisymmetric part and 6 in the symmetric part), and $j^{\lambda\mu\nu}$ due to three orthogonality conditions and antisymmetry in the last two indices has only 9 independent components. The tensor $h^{\mu \nu}$ may be parametrized in terms of the vector $w^\sigma$
\bel{eq:hmunu}
h^{\mu \nu} = \epsilon^{\mu\nu\rho\sigma} u_\rho w_\sigma,
\eel
where $w \cdot u = 0$.

%%%%%%%%%%%%%%%%%%%%%%%%%%%%%%%%%%%%%%%%%%%%%%
\subsection{Landau matching conditions} 

If the vector $u$ appearing in Eqs.~\EQn{eq:Ndec}, \EQn{eq:TdecT}, and \EQn{eq:Sdec} is identified with the hydrodynamic flow, one can immediately notice that the general forms of the tensors $N^\mu$, $T^{\mu \alpha}$, and $S^{\mu, \alpha \beta}$ contain contributions that have the structure of the equilibrium ones. The free coefficients can be fixed by the following (Landau) matching conditions
\begin{align}\label{eq:matchN}
N^\mu u_\mu &= N^\mu_{\rm eq} u_\mu, \\
T^{\mu\nu} u_\mu u_\nu &= T^{\mu\nu}_{\rm eq} u_\mu u_\nu, \label{eq:matchT} \\
S^{\lambda, \mu\nu} u_\lambda  
&= S^{\lambda, \mu\nu}_{\rm eq} u_\lambda.
\label{eq:matchS}
\end{align}
The consequences of \EQSM{eq:matchN}{eq:matchS} are straightforward:
\begin{align}\label{eq:matchNs}
a &= {\bar n}(T,\xi, k^2, \omega^2) , \\
c &= {\bar \varepsilon}(T,\xi, k^2, \omega^2), \label{eq:matchTs} \\
f^\mu &= A(T,\xi) k^\mu, \label{eq:matchSk} \\
w^{\mu} &= A_1(T,\xi) \omega^\mu.  \label{eq:matchSo}
\end{align}
A natural interpretation of the above equations is that for any $a, c, f^\mu$, and $w^{\mu}$ appearing in the general tensor decompositions \EQn{eq:Ndec}, \EQn{eq:Tdec}, and \EQn{eq:Sdec}, one can choose $T, \xi, k^\mu,$ and $\omega^\mu$  (by solving \EQSM{eq:matchNs}{eq:matchSo}) in such a way that certain parts of $N^\mu$, $T^{\mu \alpha}$, and $S^{\mu, \alpha \beta}$ have the form of the equilibrium tensors. Then, deviations from local equilibrium are defined by the formulas
\begin{align}\label{eq:devN}
\delta N^\mu  &= V^\mu, \\
\delta T^{\mu\nu}_s &= -\Pi \Delta^{\mu\nu} + W^\mu u^\nu + W^\nu u^\mu + \pi^{\mu\nu}, \label{eq:devTs} \\
\delta T^{\mu\nu}_a &= d^\mu_a u^\nu - d^\nu_a u^\mu + e^{\mu\nu}_a, 
\label{eq:devTa} \\
\delta S^{\lambda, \mu\nu} &= \Sigma^{\lambda\mu} u^\nu - \Sigma^{\lambda\nu} u^\mu + \phi^{\lambda \mu\nu} \label{eq:devS},
\end{align}
where 
\begin{align}\label{eq:V}
V^\mu &= b^\mu - n_t t^\mu, \\
\Pi &= e - {\bar P}_{k\omega},  \label{eq:Pi} \\
W^\mu &= d^\mu_s - P_t t^\mu,   \label{eq:W}  \\
\pi^{\mu\nu} &= e^{\LAB \mu\nu \RAB}_s - P_{k\omega} (\, k^{\LAB \mu} k^{\nu \RAB}  +\,\omega^{\LAB \mu} \omega^{\nu \RAB} ), \label{eq:pi} \\
\Sigma^{\lambda \mu} &= i^{\lambda \mu} - \f{A}{2} t^{\lambda \mu}, 
\label{eq:Sig} \\
\phi^{\lambda\mu\nu} &= j^{\lambda\mu\nu} - \f{A}{2} ( \Delta^{\lambda \mu} k^\nu - \Delta^{\lambda \nu} k^\mu). \label{eq:phi3} 
\end{align}
Equations~\EQSMn{eq:devN}{eq:devS} have exactly the same forms as those analyzed in~\CIT{Biswas:2023qsw}. Hence, we can directly use the results obtained in~\CITn{Biswas:2023qsw} to express the tensors appearing on the right-hand sides of~\EQSM{eq:devN}{eq:devS} by the (gradients of) hydrodynamic variables multiplied by the appropriate kinetic coefficients. Coming back to general expressions for macroscopic variables, we find that the baryon current is given by \EQ{eq:Ndec} where $a = {\bar n}(T,\xi, k^2, \omega^2)$ and
\bel{eq:b}
b^\mu = \lambda \nabla^\mu \xi + n_t t^\mu.
\eel
Here $\lambda \geq 0$ is the diffusion coefficient. The energy-momentum tensor is given by \EQ{eq:Tdec} with the coefficient  
$c = {\bar {\varepsilon}}(T,\xi, k^2, \omega^2)$ and
\begin{align}\label{eq:ds}
d^\mu_s &= -\kappa (D u^\mu - \beta \nabla^\mu T ) + P_t t^\mu, \\
d^\mu_a &=\lambda_a \beta^{-1} 
(\beta D u^\mu + \beta^2  \nabla^\mu T - 2 k^\mu), \label{eq:da} \\
e &= {\bar P} -\zeta \theta - (1/3)  
P_{k\omega} (k^2 + \omega^2), \\
e^{\LAB \mu\nu \RAB}_s &= 2 \eta \sigma^{\mu\nu} + P_{k\omega} (\, k^{\LAB \mu} k^{\nu \RAB}  +\,\omega^{\LAB \mu} \omega^{\nu \RAB} ), \\
e^{\mu \nu}_a &=  \gamma 
\beta \nabla^{[ \mu} u^{\nu ]}.
\end{align}
Here $\eta$ and $\zeta$ are the shear and bulk viscosity coefficients, respectively, $\kappa$ is the thermal conductivity, $D = u^\mu \p_\mu$ is the convective derivative, $\theta = \p_\mu u^\mu$ is the expansion scalar, $\sigma^{\mu\nu} = \p^{\LAB \mu} u^{\nu \RAB}$ is the shear flow tensor, while $\lambda_a$ and $\gamma$ are the ``new'' spin kinetic coefficients introduced for the first time in~\CITn{Hattori:2019lfp}.

In the case of the spin tensor, we first use the matching conditions \EQSTWOn{eq:matchSk}{eq:matchSo}. Then, using Eqs.~(38)--(41) from \CIT{Biswas:2023qsw}, one obtains (see also Appendix~\ref{sec:contr})
\bel{eq:i}
i^{\lambda\mu} &=& -\chi_1 \Delta^{\lambda\mu} u^\beta \nabla^\alpha \omega_{\alpha\beta}
- \chi_2 u_\nu \nabla^{\LAB \lambda} \omega^{\mu \RAB \nu} 
 -\chi_3 u_\nu \Delta^{ [ \mu}_\rho \nabla^{ \lambda ] } \omega^{\rho \nu} 
+\f{A}{2} t^{\lambda \mu}, \\
j^{\lambda\mu\nu} &=& 
\f{\chi_4 }{2} \nabla^{ \lambda} \omega^{\LAB \mu \RAB \LAB \nu \RAB} 
+\f{A}{2} ( \Delta^{\lambda \mu} k^\nu - \Delta^{\lambda \nu} k^\mu), \label{eq:j}
\eel
where $\chi_1$, $\chi_2$, $\chi_3$ and $\chi_4$ are (nonnegative) spin kinetic coefficients originally introduced in~\CIT{Biswas:2023qsw}.~\footnote{The appearance of three different kinetic coefficients in \EQn{eq:i} is related to the decomposition of the tensor $\Sigma^{\lambda\mu}$ in \EQn{eq:devS} into three parts: symmetric with nonzero trace, symmetric traceless, and antisymmetric. } 

The formulas derived above, together with the conservation laws \EQn{eq:con_eqN}, form a framework of {\it dissipative spin hydrodynamics} that can be considered an analogy to the relativistic Navier-Stokes theory. Since the latter is known to suffer from unstable behavior (in the spinless case), the current formulation most likely requires a future extension that includes second-order corrections in gradients. This can be done in a straightforward way along the guidelines given in~\CIT{Biswas:2023qsw}. Since this problem is essentially a technical one and contains many details, we leave it for a separate study. We emphasize that the arguments presented so far, and restricted to the first-order terms in gradients, already illustrate the main ideas of our work.

%%%%%%%%%%%%%%%%%%%%%%%%%%%%%%%%%%%%%%%%%%%%%%
\subsection{Systematics of expansion} 

Our expressions for the baryon current~\EQn{eq:NmuG}, the energy-momentum tensor~\EQn{eq:Tmunu_decomposed}, and the spin tensor~\EQn{eq:spint_GLW} were based on the expansion in the spin polarization tensor, $\omega_{\alpha\beta} = \Omega_{\alpha\beta}/T$, which in natural units is a dimensionless quantity, similarly as the ratio $\xi = \mu/T$. Hence, the expansion in $k$ and $\omega$ is well defined as controlled by dimensionless parameters. On the other hand, inclusion of dissipation introduces gradient terms, whose importance is quantified by the  Knudsen number. 

Since some of the dissipative currents are determined by a combination of the gradient terms and the spin polarization components, many works consider $\omega_{\alpha\beta}$ to be a quantity of the same order as the gradient terms (see \CIT{Hattori:2019lfp} and works based on that paper). This implies that the gradients of $\omega_{\alpha\beta}$ appearing on the right-hand sides of \EQSTWO{eq:i}{eq:j} are considered to be second-order corrections and neglected. This also directly leads to the conclusion that the terms of the form $u_\lambda S^{\lambda, \alpha\beta} \omega_{\alpha\beta}$ should be at least of the second order and neglected. However, the latter conclusion is usually ignored by stating that the spin density tensor $S^{\alpha\beta} = u_\lambda S^{\lambda, \alpha\beta}$ is of the zeroth order in gradients. Clearly, the assumption that the magnitude of $\omega_{\alpha\beta}$ is fixed by the magnitude of gradients cannot hold in general~\footnote{For example, let us consider the dimensionless expression $\beta D u^\mu - \beta^2  \nabla^\mu T - 2 k^\mu$ in~\EQ{eq:da}. For boost-invariant flows $D u^\mu = 0$ and $\nabla^\mu T = 0$, however, $k^\mu$ may be noticeably different from zero.}. Our approach suggests that the gradient corrections and the corrections arising due to the appearance of the spin polarization tensor components should be treated independently. The paper~\CITn{Florkowski:2024bfw} shows that an expansion in $\omega_{\alpha\beta}$ to second-order terms is crucial for obtaining a nontrivial and consistent treatment of thermodynamic relations. In this work we show how to extend~\CITn{Florkowski:2024bfw} by additions of dissipative (gradient) terms. 

%%%%%%%%%%%%%%%%%%%%%%%%%%%%%%%%%%%%%%%%%%%%%%
\section{Summary and conclusions} 
\label{sec:summary}

In this work we have introduced a unified framework of spin hydrodynamics that combines the results of kinetic theory for particles with spin $\onehalf$ with the IS approach for including nonequilibrium processes. The latter is used to incorporate what is usually referred to as the first-order terms (in gradients). The inclusion of the second-order terms is straightforward but quite lengthy, so we leave it as a separate project. The framework of the kinetic theory has been used to define the perfect-fluid description. Interestingly, in the presence of spin degrees of freedom, it includes terms whose form is usually attributed to dissipation. The genuine dissipative terms appear from the condition of positive entropy production that forms the basis of the IS method. They are responsible for transfers between the spin and orbital parts of the total angular momentum. 

The proposed framework solves long-standing problems encountered in previous studies based on the positive-entropy production principle (inconsistent expansion in the spin polarization tensor, problems with thermodynamic identities) and local kinetic theory (neglecting the spin-orbit coupling). It also seems to be straightforward to implement in practical applications/codes, as it circumvents possible technical difficulties connected with the use of the nonlocal collision kernel.

%%%%%%%%%%%%%%%%%%%%%%%%%%%%%%%%%%%%%%%%%%%%%%
\noindent
{\it Acknowledgments --} We thank Samapan Bhadury, Valeriya Mykhaylova,  and Radoslaw Ryblewski for clarifying discussions. This work was supported in part by the Polish National Science Centre Grant No. 2022/47/B/ST2/01372.

\appendix
\section{\texorpdfstring{Construction of the tensors $Z^{\pm \alpha \beta \ldots}_n$}{The Z tensors}} %"\texorpdfstring" is required to get rid of a LATEX warning during compilation
\label{sec:Ztensors}

The tensors  $Z^{\pm \alpha \beta \ldots}_n \equiv \int \dd P p^\alpha p^\beta \cdots f_n^\pm$ that appear in calculations involving the FD distribution functions have the same
symmetries (i.e., dependence on the product $p \cdot u$ and symmetry under exchange of any Lorentz indices) as the tensors $Z^{\alpha\beta \ldots} \equiv \int \dd P p^\alpha p^\beta \cdots e^{-\beta \cdot p}$ that appear in the classical (Boltzmann) approach, and therefore they admit the same 
decomposition in terms of generic
tensors of the same symmetry built out of the hydrodynamic flow vector $u^\alpha$ and the metric tensor $g^{\alpha\beta}$ (see Eqs.~(5.225)--(5.228) in \CIT{Cercignani:2002rh}). 

The general method for determining the functional form of the coefficients that appear in such a decomposition is to calculate appropriate traces and/or tensor contractions with the flow $u_\alpha$, so as to obtain a set of linear scalar equations, which can be easily solved. In the FD case, after a suitable variable change, this directly leads to the coefficients expressed as combinations of integrals of the  following two types~\CITn{Cercignani:2002rh}
\bel{eq:Jnm_In}
J_{n m}(\xi,z) = \int_0^\infty \f{\sinh^n y \cosh^m y }{e^{-\xi + z \cosh y}-\epsilon} \dd y, 
 \quad I_n(\xi,z) =\int_0^\infty \f{ \cosh{(ny)} }{ e^{z \cosh{y} } - \epsilon } \dd y.
\eel
The parameter $\epsilon$ introduced here is equal to $-1$ for the FD statistics. 

Within the coefficients, the functions $J_{n m}(\xi, z)$ always appear with even $n$ and can be expressed as combinations of the functions $I_{n}(\xi, z)$ by using the hyperbolic identity $\sinh^2 y = \cosh^2 y -1$ and replacing powers of $\cosh y$ by hyperbolic cosine of a multiplied argument. The relations we need are the following (for brevity of notation we shall often suppress the arguments of the functions):
\begin{align}\begin{split}\label{eq:J_I_2}
J_{20} = \f{1}{2}( I_2 - I_0), \quad
J_{21} = \f{1}{4}( I_3 - I_1), \quad
J_{22} = \f{1}{8}( I_4 - I_0 ), \quad \\
J_{23} = \f{1}{16}( I_5 + I_3 - 2I_1), \quad
J_{24} = \f{1}{32}(I_6 + 2 I_4 - I_2 -2 I_0). \quad
\end{split}\end{align}
By taking the formal limit $\epsilon \to 0$, in which
\bel{eq:ItoK}
\lim_{\epsilon \rightarrow 0} I_{n}(\pm \xi,z) \rightarrow e^{\pm \xi} K_{n}(z),
\eel
we can reproduce the results for the Boltzmann case, \EQ{eq:ZFDZB}, where the coefficients can be expressed in terms of modified Bessel functions. The latter obey a recursive relation,
\bel{eq:Knrec}
K_{n+1}(z)=K_{n-1}(z) + \f{2n}{z}K_{n}(z),
\eel
which we shall use in the intermediate steps of the calculations.

It is also important to note that it suffices to give the expressions for the tensors $Z^{\pm \alpha \beta \ldots}_0$, as the terms with $n > 0$ can be obtained by differentiation with respect to $\xi$.
\bea
Z^{\pm \alpha \beta \ldots}_n = (\pm)^n \f{\p ^n}{\p ^n \xi} Z^{\pm \alpha \beta \ldots}_0, \qquad n=1,2.
\eea

%%%%%%%%%%%%%%%%%%%%%%%%%%%%%%%%%%%%%%%%%%%%%%%%%%
\subsection*{1.~Rank 1}

Because of the Lorentz covariance, the tensor $Z^{\pm \mu}_{0}$ has to be proportional to the hydrodynamic flow vector,
\bel{A1}
Z^{\pm \alpha}_{0}=\int \dd P\, p^\alpha f^\pm_0 \equiv \mathcal{A}_1^\pm u^{\alpha}.
\eel
To explicitly compute the coefficient $\mathcal{A}_{1}^\pm$, we contract this equation with $u_\alpha$ to obtain
\bea
\mathcal{A}_{1}^{\pm} = \int \dd P  \, p \cdot u \, f^\pm_0.
\eea
In the fluid rest frame, $u = (1,0,0,0)$ and $p \cdot u = E_p$, so we can write
\bea
\mathcal{A}_{1}^{\pm } = \int \f{\dd^3 p}{(2 \pi)^3 }  f^\pm_0 = \f{1}{2 \pi^2 }\int |p|^2 \dd |p| \f{1}{e^{ \mp \xi + \f{E_p z}{m}} - \epsilon}.
\eea
By introducing the rapidity $y$, $E_p = m \cosh y$, $|p| = m \sinh y$, $\dd |p| = m \cosh y \dd y$, and changing the integration variable, we find that
\bel{eq:K02mu}
\mathcal{A}_{1}^{\pm } = \frac{m^3}{2\pi^2} \int_0^\infty   \f{\sinh ^2 y \cosh y }{e^{ \mp \xi + z\cosh y} - \epsilon} \, \dd y = \f{m^3}{2\pi^2}J_{21}(\pm \xi, z) = \f{m^3}{8\pi^2}
\LSB I_{3}(\pm \xi, z) - I_{1}(\pm \xi, z) \RSB,
\eel
where we used the definitions~\EQn{eq:Jnm_In}. Now we can check the Boltzmann limit $\epsilon \to 0$,
\begin{align} \label{eq:calK2toK2}
\begin{split}
\lim_{\epsilon \rightarrow 0 } \, \mathcal{A}_{1}^{\pm} 
= \f{m^3}{8 \pi^2} e^{\pm \xi} \LSB K_{3}(z)-K_{1}(z) \RSB 
=  \f{m^2 T }{2 \pi^2} e^{\pm \xi} K_{2}(z),
\end{split}
\end{align}
where we used~\EQ{eq:ItoK} and the recursive relation~\EQn{eq:Knrec}. Thus, indeed, the classical limit given by~\EQn{eq:Zmu} is reproduced.
Higher-order corrections in the spin polarization tensor can be obtained by differentiation of \EQ{eq:K02mu} with respect to $\xi$.
\bea
Z_1^{\pm \alpha} = \pm  u^\alpha \f{\p}{\p \xi} \mathcal{A}_{1}^{\pm}, \\
Z_2^{\pm \alpha} = u^\alpha \f{\p^2}{\p \xi^2} \mathcal{A}_{1}^\pm.
\eea

%%%%%%%%%%%%%%%%%%%%%%%%%%%%%%%%%%%%%%%%%%%%%%%%
\subsection*{2.~Rank 2}

For the rank-2 symmetric tensor, we use the following decomposition:
\bea
Z^{\pm \alpha \beta}_0 = \int \dd P p^\alpha p^\beta f^\pm_0 = \mathcal{A}_{2}^\pm g^{\alpha \beta} + \mathcal{B}_{2}^\pm u^\alpha u^\beta.
\eea
By calculating the contraction $Z^{\pm \alpha \beta}_0  u_\alpha u_\beta$ and the trace $Z^{\pm \alpha}_{0 \,\,\,\alpha}$, we obtain a system of two equations for $\mathcal{A}^\pm_2$ and $\mathcal{B}^\pm_2$ (once again, we consider the local rest frame and use rapidity as the integration variable)
\bea
\int \dd P (p \cdot u)^2 f^\pm_0 = \f{1}{2\pi^2}\int_0^\infty  |p|^2 \,  E_{p} \,  f_{0}^\pm \, \dd |p| = \f{m^3}{2\pi^2} J_{22}(\pm \xi, z) = \mathcal{A}_2^\pm + \mathcal{B}_2^\pm,\\
\int \dd P p^2 f^\pm_0 = \f{1}{2 \pi^2} \int_0^\infty \f{|p^2|}{E_p}m^2 f_{0}^\pm d|p| = \f{m^4}{2 \pi^2} J_{20}(\pm \xi, z) = 4\mathcal{A}_2^\pm + \mathcal{B}_2^\pm.
\eea
It is solved by
\bea
\mathcal{A}_2^\pm = \f{m^4}{3 \cdot 2 \pi^2} (J_{20} - J_{22}) = - \f{m^4}{8 \cdot 3 \cdot 2 \pi^2} (I_4 - 4 I_2 + 3I_0),\\
\mathcal{B}_2^\pm = \f{m^4}{3 \cdot 2 \pi^2} (4J_{22} - J_{20}) = \f{m^4}{2 \cdot 3 \cdot 2 \pi^2} (I_4 - I_2).
\eea
In the classical limit, we obtain
\bea
\lim_{\epsilon \rightarrow 0} \mathcal{A}_2^\pm =
 -\f{ m^4 }{ 48 \pi^2 } e^{\pm \xi} 
\LB K_{4}-4K_{2}+3K_{0} \RB 
= - \f{m^2 T^2}{2 \pi^2} e^{\pm \xi} K_2(z),\\
\lim_{\epsilon \rightarrow 0} \mathcal{B}_2^\pm = 
\f{m^4}{12\pi^2} e^{\pm \xi} \LB K_4 - K_2 \RB =  \f{m^3T}{2\pi^2} e^{\pm \xi}  K_3(z) .
\eea
Higher orders in spin are given by 
\begin{align}
\begin{split}
Z_{1}^{\pm \alpha \beta} = \pm u^\alpha u^\beta \f{\p}{\p \xi} \mathcal{A}_{2}^\pm \pm g^{\alpha \beta} \f{\p}{\p \xi} \mathcal{B}_{2}^{\pm},\\
Z_{2}^{\pm \alpha \beta} = u^\alpha u^\beta \f{\p^2}{\p \xi^2} \mathcal{A}_{2}^\pm + g^{\alpha \beta} \f{\p^2}{\p \xi^2} \mathcal{B}_{2}^{\pm}.
\end{split}
\end{align}

%%%%%%%%%%%%%%%%%%%%%%%%%%%%%%%%%%%%%%%%%%%%%%%%%%%%%%%%%
\subsection*{3.~Rank 3}

For the rank-3 symmetric tensor, we use the following decomposition:
\bea
Z^{\pm \alpha \beta \gamma}_0 = \int \dd P \ p^\alpha p^\beta p^\gamma f^\pm_0 = \mathcal{A}_3^\pm (g^{\alpha \beta} u^\gamma + g^{\alpha \gamma} u^\beta + g^{\gamma \beta} u^\alpha) + \mathcal{B}_3^\pm u^\alpha u^\beta u^\gamma.
\eea
Calculating the contraction $Z_0^{\pm \alpha \beta \gamma} u_\alpha u_\beta u_\gamma$ and the partial trace with contraction $Z_0^{\pm \alpha}{}_\alpha{}^\gamma u_\gamma$ leads to the following set of two equations:
\bea
\int \dd P (p \cdot u)^3 f^\pm_0 = \f{1}{2 \pi^2 }\int_0^\infty |p|^2 \, E_p^2 \, f_0^\pm \, \dd |p|
= \f{m^5}{2 \pi^2} J_{23}(\pm \xi, z) = 3\mathcal{A}_3^\pm + \mathcal{B}_3^\pm,
\eea
\bea
\int \dd P p^2 (p\cdot u) f^\pm_0 = \f{1}{2 \pi^2 } \int_0^\infty |p|^2 m^2 f_0^\pm \dd |p|
= \f{m^5}{2 \pi^2} J_{21}(\pm \xi, z) = 6\mathcal{A}_3^\pm +\mathcal{B}_3^\pm,
\eea
which are solved by \footnote{It can also be noted that $J_{21}-J_{23}=-J_{41}.$}
\begin{equation}
\mathcal{A}_3^\pm = \f{m^5}{3 \cdot 2\pi^2} (J_{21} - J_{23}) =
\f{m^5}{16 \cdot 3 \cdot 2 \pi^2} (-I_5 + 3 I_3 -2 I_1),
\end{equation}
\bea
\mathcal{B}_3^\pm = \f{m^5}{2\pi^2} (2 J_{23} - J_{21}) = \f{m^5}{8 \cdot 2\pi^2} (I_5 - I_3),
\eea
with the classical limit
\bea
\lim_{\epsilon \rightarrow 0} \mathcal{A}_3^\pm  = -\f{m^5}{16 \cdot 6 \pi^2} e^{\pm \xi}\LB K_{5}-3K_{3}+2K_{1} \RB = - \f{m^3 T^2}{2 \pi^2}  e^{\pm \xi} K_3(z),
\eea
\bea
\lim_{\epsilon \rightarrow 0} \mathcal{B}_1^\pm 
= \f{m^5}{8 \cdot 2 \pi^2} e^{\pm \xi} \LB K_5 - K_3 \RB \ 
= \f{m^4 T}{2 \pi^2} e^{\pm \xi} K_4(z).
\eea
Higher spin corrections are equal to
\begin{align}
\begin{split}
Z_{1}^{\pm \alpha \beta \gamma} =  \pm( g^{ \alpha \beta} u^\gamma + g^{\alpha \gamma} u^\beta + g^{\gamma \beta} u^\alpha) \f{\p}{\p \xi} \mathcal{A}_{3}^\pm \pm u^\alpha u^\beta u^\gamma  \f{\p}{\p \xi} \mathcal{B}_{3}^{\pm}, \\
Z_{2}^{\pm \alpha \beta \gamma} = (g^{\alpha \beta} u^\gamma + g^{\alpha \gamma} u^\beta + g^{\gamma \beta} u^\alpha)\f{\p^2}{\p \xi^2} \mathcal{A}_{3}^\pm + u^\alpha u^\beta u^\gamma \f{\p^2}{\p \xi^2} \mathcal{B}_{3}^{\pm}.
\end{split}
\end{align}

%%%%%%%%%%%%%%%%%%%%%%%%%%%%%%%%%%%%%%%%%%%%%%%%%%%%%
\subsection*{4.~Rank 4}

For the rank-4 symmetric tensor, we use the decomposition
\begin{align}\label{eq:rank4}
\begin{split}
Z^{\pm \alpha \beta \gamma \delta}_0 &= \int \dd P \ p^\alpha p^\beta p^\gamma p^\delta f^\pm_0 = \mathcal{A}_{4}^\pm (g^{\alpha \beta} g^{\gamma \delta} + g^{\alpha \gamma} g^{\beta \delta} + g^{\gamma \beta} g^{\alpha \delta})\\
&+ \mathcal{B}^\pm_4 (g^{\alpha \beta} u^\gamma u^\delta + g^{\alpha \gamma} u^\beta u^\delta
+ g^{\gamma \beta} u^\alpha u^\delta + g^{\alpha \delta} u^\gamma u^\beta + g^{\delta \gamma} u^\beta u^\alpha + g^{\delta \beta} u^\alpha u^\gamma)\\
&+ \mathcal{C}_4^\pm u^\alpha u^\beta u^\gamma u^\delta.
\end{split}
\end{align}
Calculating the contraction $Z_0^{\pm \alpha \beta \gamma \delta} u_\alpha u_\beta u_\gamma u_\delta$, the contraction and partial trace $Z_0^{\pm \alpha}{}_\alpha{}^{\gamma \delta} u_\gamma u_\delta$, as well as the trace $Z_0^{\pm \alpha}{}_\alpha{}^\gamma{}_\gamma$ leads to the following set of three linear equations for the scalar coefficients
\begin{align}\begin{split}
&\int \dd P (p \cdot u)^4 f^\pm_0= \f{1}{ 2\pi^2 } \int_0^\infty |p|^2 \, E_p^3 \, f_0^\pm \, \dd |p| \\
&= \f{m^6}{2\pi^2} J_{24} (\pm \xi, z)
= 3\mathcal{A}_4^\pm + 6\mathcal{B}_4^\pm +\mathcal{C}_4^\pm,
\end{split}\end{align}
\begin{align}\begin{split}
&\int \dd P p^2 (p \cdot u)^2 f^\pm_0 = \f{1}{ 2\pi^2 } \int_0^\infty |p|^2 \, E_p \, m^2 \, f_0^\pm \, \dd |p| \\
&= \f{m^6}{2\pi^2}J_{22} (\pm \xi, z)
= 6\mathcal{A}_4^\pm + 9 \mathcal{B}_4^\pm + \mathcal{C}_4^\pm,
\end{split}\end{align}
\begin{align}\begin{split}
&\int \dd P (p^2)^2 f^\pm_0= \f{1}{ 2\pi^2 } \int_0^\infty \f{|p|^2}{E_p} m^4 \, f_0^\pm \, \dd |p|\\
&= \f{m^6}{2\pi^2}J_{20} (\pm \xi, z)
=24\mathcal{A}_4^\pm + 12 \mathcal{B}_4^\pm + \mathcal{C}_4^\pm,
\end{split}\end{align}
which are solved by~\footnote{Note that $J_{24}-2J_{22}+J_{20} = J_{60}.$}
\begin{equation}
\mathcal{A}_4^\pm = \f{m^6}{15 \cdot 2 \pi^2}  (J_{24}-2J_{22}+J_{20})
= \f{m^6}{32\cdot 15 \cdot 2 \pi^2}\LB I_6 -6I_4 +15I_2 - 10I_0 \RB,
\end{equation}
\begin{align}\begin{split}
\mathcal{B}_4^\pm &=  -\f{m^6}{15 \cdot 2 \pi^2}  \LB 6J_{24} -7J_{22} +J_{20}  \RB = -\f{m^6}{32 \cdot15 \cdot 2 \pi^2} \LB 6I_6 -16 I_4 + 10I_2 \RB,
\end{split}\end{align}
\begin{align}\begin{split}
\mathcal{C}_4^\pm &= \f{m^6}{5 \cdot 2 \pi^2} \LB 16 J_{24} -12  J_{22} + J_{20} \RB  =\f{m^6}{5 \cdot 2 \cdot 2 \pi^2} \LB I_6 - I_4 \RB.
\end{split}\end{align}
Again, in the classical limit, we obtain the same result as in
\cite{Cercignani:2002rh},
\begin{align}\begin{split}
\lim_{\epsilon \rightarrow 0} \mathcal{A}_4^\pm = \f{m^6}{32\cdot 15 \cdot 2 \pi^2} e^{\pm \xi} \LB K_6 -6K_4 +15K_2 - 10K_0 \RB = \f{m^3 T^3}{2 \pi^2} e^{\pm \xi} K_3(z),
\end{split}\end{align}
\begin{align}\begin{split}
\lim_{\epsilon \rightarrow 0} \mathcal{B}_4^\pm = \f{m^6}{16\cdot 15 \cdot 2 \pi^2} e^{\pm \xi} \LB -3 K_6 + 8 K_4 - 5 K_2 \RB = -\f{m^4 T^2 }{2 \pi^2 } e^{\pm \xi} K_4(z),
\end{split}\end{align}
\begin{align}\begin{split}
\lim_{\epsilon \rightarrow 0} \mathcal{C}_4^\pm = \f{m^6}{ 5 \cdot 2 \cdot 2 \pi^2} e^{\pm \xi} \LB K_6 - K_4 \RB = \f{m^5 T }{2 \pi^2} e^{\pm \xi} K_5(z).
\end{split}\end{align}
Higher spin corrections $Z_k^{\pm \alpha \beta \gamma \delta}$ have the same form as~\EQn{eq:rank4} but with coefficients $\mathcal{A}_4^\pm$, $\mathcal{B}_4^\pm$, $\mathcal{C}_4^\pm$ replaced by
\bea
\mathcal{A}_{4n}^\pm = (\pm 1)^n \f{\p^n}{\p^n \xi} \mathcal{A}_{4}^\pm, \quad \mathcal{B}_{4n}^\pm = (\pm 1)^n \f{\p^n}{\p^n \xi} \mathcal{B}_{4}^\pm, \quad \mathcal{C}_{4n}^\pm = (\pm 1)^n \f{\p^n}{\p^n \xi} \mathcal{C}_{4}^\pm. \quad
\eea

%%%%%%%%%%%%%%%%%%%%%%%%%%%%%%%%%%%%%%%%%%%%%%%%
\section{Spin degrees of freedom and tensor contractions}
In the spin configuration space, one can introduce the following measure~\CITn{Florkowski:2018fap},
\bel{eq:dSn}
\dd S = \f{m}{\pi \spin}  \, \dd ^4 s \, \delta(s \cdot s + \spin^2) \, \delta(p \cdot s).
\eel
The two delta functions control here the normalization of the spin vector and its orthogonality to particle momentum. The prefactor in \EQn{eq:dS} is chosen such as to yield the normalization condition
\bel{eq:dS2}
\int \dd S = 2,
\eel
which reflects two possible orientations of the spin $\onehalf$. Further useful integrals include~\CITn{Florkowski:2018fap}
\bel{eq:dSs} 
\int \dd S \,s_{\alpha} &=& 0,  \\
\int \dd S \,s_\sigma s_\rho &=& 
-\f{2 \spin^2 }{3}\LB g_{\sigma \rho} + p_\sigma p_\rho \RB. \label{eq:dSss}
\eel
They can be used to derive three other useful integrals over the spin configuration space
\bel{eq:dSos} 
\int \dd S \, \omega : s &=& 0,  \\
\int \dd S \, s^{\mu\nu} \, \omega : s &=&    \f{4 \spin^2 }{3 m^2}
\LSB m^2  \omega^{\mu \nu} + p_\alpha \LB p^\mu \omega^{\nu\alpha}
-p^\nu \omega^{\mu\alpha} \RB \RSB
,  \label{eq:dSsos} \\
\int \dd S \, (\omega : s)^2  &=& 
\f{4 \spin^2 }{3 m^2}
\LB m^2  \omega : \omega + 2 \, p^\alpha p^\beta \omega^{\gamma}_{\,\,\, \alpha} \omega_{\beta\gamma} \RB. \label{eq:dSosos}
\eel
In calculations of the baryon current, the energy-momentum tensor, and the spin tensor, these expressions multiply other terms under the integral over the momentum space. Then, the next step is the integration over $\dd P$ (see $Z^{\alpha \beta...}$ in Appendix A), and then the resulting expressions involve certain contractions with $\omega^{\nu \alpha}$ or $\omega^{\gamma}_{\phantom{\gamma}\alpha} \omega_{\beta \gamma}$, which we present here.

Computing the baryon current $N^{\mu}_{eq}$ requires the contraction $Z^{\mu \alpha \beta} \omega^\gamma_{\phantom{\gamma}\alpha} \omega_{\beta \gamma}$
which can be expressed as
\bel{eq:Z3oo}
Z^{\mu \alpha \beta} \omega^\gamma_{\phantom{\gamma}\alpha} \omega_{\beta \gamma} = -\f{T^5}{2\pi^2} z^3 \LSB K_3(z) R^\mu_{N1} - z K_4(z) R^\mu_{N2} \RSB,
\eel 
with
\begin{align}\begin{split}\label{eq:RN1}
R^\mu_{N1} &= \LB g^{\mu\alpha} u^{\beta} + g^{\mu\beta} u^{\alpha} + g^{\beta\alpha} u^{\mu} \RB \omega^\gamma_{\phantom{\gamma}\alpha} \omega_{\beta \gamma} = \LB 2 \omega^2 - 4 k^2 \RB u^\mu + 2 t^\mu,\\
R^\mu_{N2} &= u^\mu u^\alpha u^\beta \omega^\gamma_{\phantom{\gamma}\alpha} \omega_{\beta \gamma} =- k^2  u^\mu.
\end{split}\end{align}

The energy-momentum tensor $T^{\mu \nu}_{eq}$ involves the contraction $Z^{\mu\nu\alpha\beta} \omega^\gamma_{\phantom{\gamma}\alpha} \omega_{\beta \gamma}$, which is equal to
\bel{eq:Z4oo}
Z^{\mu\nu\alpha\beta} \omega^\gamma_{\phantom{\gamma}\alpha} \omega_{\beta \gamma} &=& \f{T^6}{2\pi^2} z^3  \LSB 
K_3(z) R_{T1}^{\mu\nu} - z K_4(z) R_{T2}^{\mu\nu} + z^2 K_5(z) R_{T3}^{\mu\nu}
\RSB
\eel
with
\begin{align}\begin{split}\label{eq:RT1}
R_{T1}^{\mu\nu} &= \LB g^{\mu\nu} g^{\alpha\beta} + g^{\mu\alpha} g^{\nu\beta} + g^{\alpha\nu} g^{\mu\beta} \RB \omega^{\gamma}_{\phantom{\gamma}\alpha} 
\omega_{\beta \gamma} \\
&= 2 \LSB
g^{\mu\nu} (2\omega^2-k^2)-u^\mu u^\nu (k^2+\omega^2) - (k^\mu k^\nu +\omega^\mu \omega^\nu) + u^\mu t^\nu + u^\nu t^\mu \RSB, \\ 
R_{T2}^{\mu\nu} &= \LB g^{\mu\nu} u^{\alpha} u^{\beta} +  g^{\mu\alpha} u^{\nu} u^{\beta} + g^{\alpha\nu} u^{\mu} u^{\beta} + g^{\mu\beta} u^{\alpha} u^{\nu} + g^{\beta\alpha} u^{\nu} u^{\mu} + g^{\beta\nu} u^{\mu} u^{\alpha} \RB \omega^{\gamma}_{\phantom{\gamma}\alpha} 
\omega_{\beta \gamma} \\
&= -k^2 g^{\mu\nu} + 2 (\omega^2 - 3 k^2) u^\mu u^\nu + 2 (u^\mu t^\nu + u^\nu t^\mu),\\
R_{T3}^{\mu\nu} &= u^{\mu} u^{\nu} u^{\alpha} u^{\beta}  \omega^{\gamma}_{\phantom{\gamma}\alpha} 
\omega_{\beta \gamma} = -k^2 u^\mu u^\nu,
\end{split}\end{align}
Finally, for the spin tensor $S^{\lambda, \mu \nu}_{eq}$ we need the following contraction
\begin{align}\begin{split}\label{eq:ZZ}
& Z^{\lambda \alpha \mu}\omega^{\nu}_{\phantom{\nu}\alpha} - Z^{\lambda \alpha \nu}\omega^{\mu}_{\phantom{\mu}\alpha} \\
& = -\f{T^5}{2\pi^2} z^3 \LSB K_3(z) \LB g^{\lambda\alpha} u^{\mu} + g^{\lambda\mu} u^{\alpha} + g^{\mu\alpha} u^{\lambda} \RB - z K_4(z) u^\lambda u^\alpha u^\mu \RSB \omega^\nu_{\,\,\,\alpha}\\
& \,\,\,\, + \f{T^5}{2\pi^2} z^3 \LSB K_3(z) \LB g^{\lambda\alpha} u^{\nu} + g^{\lambda\nu} u^{\alpha} + g^{\nu\alpha} u^{\lambda} \RB - z K_4(z) u^\lambda u^\alpha u^\nu \RSB \omega^\mu_{\,\,\,\alpha} \\
& = -\f{T^5}{2\pi^2} z^3 \LSB K_3(z) \LB 
t^{\lambda\mu\nu}
+ u^{\lambda} \omega^{\nu\mu} - u^{\lambda} \omega^{\mu\nu} \RB  
- z K_4(z) u^\lambda \LB u^\mu k^\nu - u^\nu k^\mu \RB \RSB.
\end{split}\end{align}

\section{Pressure for the FD gas}
\label{sec:pressFD}

Let us show that in traditional hydrodynamics the current $\mathcal{N}^\mu$ defined by \EQ{eq:CNmu} is equal to the product of local pressure and the hydrodynamic flow divided by temperature,
\bel{eq:mathcalN}
\mathcal{N}^\mu = P \beta u^\mu  \equiv P \beta^\mu.
\eel
Omitting spin and the antiparticle part of \EQn{eq:CNmu}, we can write 
\bea
u_\mu \mathcal{N}^\mu = -\int dP p^\mu u_\mu \ln(1-f) = 
- \frac{1}{2 \pi^2} \int_0^\infty \frac{p^2 \dd p}{E_p} \ln(1-f) E_p,
\eea
where we chose the local rest frame (LRF), in which $p=(m,0,0,0)$ and $p^\mu u_\mu = E_p$. Now we perform integration by parts,
\bel{eq:parts}
- \int_0^\infty p^2 \dd p \ln(1-f) = - \f{p^3}{3} \ln(1-f) \Bigg|_{p=0}^\infty - \int_0^\infty \f{p^3 \dd p}{3} \f{1}{1-f}
\f{\dd f}{\dd E_p} u_p,
\eel
where we have introduced the notation $u_p =\dd E_p/\dd p$. Since the boundary term vanishes, we are left with the second term only, which gives
\bel{eq:uN1}
u_\mu \mathcal{N}^\mu = \f{1}{6 \pi^2} \int_0^\infty \f{\beta u_p \dd p}{e^{-\xi + E_p\beta} + 1} = \f{\beta}{6 \pi^2} \int_0^\infty f \, p^3 u_p \dd p.
\eel
On the other hand, the grand potential $\Omega_G = -PV$ of a Fermi gas is given by a sum over all quantum states~\CITn{Landau:1980mil}
\bel{eq:Og}
\Omega_G = -T \sum_p \ln \LSB 1 +\exp{\LB \f{\mu - E_p}{T} \RB} \RSB = -T \f{V}{2\pi^2} \int_0^\infty p^2 \dd p \ln \LSB 1 + \exp (\xi - E_p \beta) \RSB.
\eel
Comparing Eqs.~\EQn{eq:uN1} and \EQn{eq:Og}, we reproduce \EQ{eq:mathcalN}.
\section{Contraction of rank-3 tensors antisymmetric in two indices}
\label{sec:contr}
As per \EQ{eq:Sdec}, the spin tensor can be parametrized in terms of simpler vectors and tensors in the following way:
\bea
S^{\lambda, \mu\nu} &=& u^\lambda \LSB \LB f^\mu u^\nu - f^\nu u^\mu \RB
+ h^{\mu \nu} \RSB + i^{\lambda \mu} u^\nu - i^{\lambda \nu} u^\mu
+ j^{\lambda\mu\nu},
\eea
with the simplifying constraints
\begin{align}\begin{split}\label{eq:constr}
f^\mu u_\mu = 0, \quad h^{\mu \nu} = - h^{\nu \mu}, \quad h^{\mu \nu} u_\mu = 0, \quad h^{\mu \nu} = \epsilon^{\mu\nu\rho\sigma} u_\rho w_\sigma, \ \text{where} \ w \cdot u = 0,\\ i^{\lambda \mu} u_\lambda = i^{\lambda \mu} u_\mu = 0, \quad j^{\lambda\mu\nu} = - j^{\lambda\nu\mu}, \quad j^{\lambda\mu\nu} u_\lambda = j^{\lambda\mu\nu} u_\mu = j^{\lambda\mu\nu} u_\nu = 0.
\end{split}\end{align}
Now, let us compute the contraction of two such tensors.
\begin{align}
\begin{split}
S^{\lambda, \mu\nu}S'_{\lambda, \mu\nu} &= \big[ u^\lambda \LB \LB f^\mu u^\nu - f^\nu u^\mu \RB
+ h^{\mu \nu} \RB + i^{\lambda \mu} u^\nu - i^{\lambda \nu} u^\mu \\
&\quad+ j^{\lambda\mu\nu} \big]
\LSB u_\lambda \LB \LB f'_\mu u_\nu - f'_\nu u_\mu \RB
+ h'_{\mu \nu} \RB + i'_{\lambda \mu} u_\nu - i'_{\lambda \nu} u_\mu
+ j'_{\lambda\mu\nu} \RSB \\
&= u^\lambda u_\lambda [2 f^\mu f'_\mu u^\nu u_\nu
- 2 f^\mu u_\mu \ f'_\nu u^\nu
- 2 h^{\mu \nu} u_\mu \ f'_\nu
- 2 h'_{\mu \nu} u^\mu \ f^\nu 
+ h^{\mu \nu} h'_{\mu \nu} ] \\
&\quad+ \LB i^{\lambda \mu} (...) 
- i ^{\lambda \nu} (...) \RB u_\lambda
+ \LB i'_{\lambda \mu}  (...) 
- i'_{\lambda \nu} (...) \RB u^\lambda 
- 2i^{\lambda \nu} u_\nu \ i'_{\lambda \mu} u^\mu\\
&\quad+ j^{\lambda \mu \nu} \LB u_\lambda (...)+ u_\nu (...)- u_\mu (...) \RB
+ j'_{\lambda \mu \nu} \LB u^\lambda (...) + u^\nu (...) - u^\mu (...) \RB \\
&\quad+ 2 i^{\lambda \mu} i'_{\lambda \mu} \ u^\nu u_\nu
+ j^{\lambda \mu \nu} j'_{\lambda \mu \nu}\\
&= 2f^\mu f'_\mu + h^{\mu \nu} h'_{\mu \nu} + 2 i^{\mu \nu} i'_{\mu \nu} + j^{\lambda \mu \nu} j'_{\lambda \mu \nu}.
\end{split}
\end{align}
Constraints ~\EQn{eq:constr} were used to eliminate most of the terms. Using the definition of $h^{\mu \nu}$ leads to further simplification,
\bea
h^{\mu \nu} h'_{\mu \nu} &= \epsilon^{\mu \nu \rho \sigma} \epsilon_{\mu \nu \alpha \beta} u_\rho w_\sigma u^\alpha w'^\beta = -2 (g^{\rho}_\alpha g^{\sigma}_\beta - g^{\rho}_\beta g^{\sigma}_\alpha) u_\rho w_\sigma u^\alpha w'^\beta = -2 w_\mu w'^\mu,
\eea
whereas the tensor $i$ can be split into the antisymmetric and the symmetric part, in the latter of which, in turn, we can isolate a part with nonzero trace and a traceless part $i^{\langle \mu \nu \rangle}$,
\beq
i^{\mu \nu} = i^{[ \mu \nu ]} + i^{( \mu \nu )} =  i^{[ \mu \nu ]} - \tilde{i} \Delta^{\mu \nu} + i^{\langle \mu \nu \rangle},
\eeq
where $\tilde{i} = -(1/3) i^{( \mu}{}_{\mu )}$ and $\Delta^{\mu \nu} = g^{\mu \nu} -u^\mu u^\nu$. A contraction of a symmetric and an antisymmetric tensor is equal to zero, as is
\begin{align}
\Delta^{\mu \nu} i_{\langle \mu \nu \rangle}' = \Delta^{\mu \nu} \LB i'_{(\mu \nu)} + \tilde{i'}\Delta_{\mu \nu} \RB = 
g^{\mu \nu} i'_{(\mu \nu)} - u^\mu u^\nu i'_{(\mu \nu)} + 3 \tilde{i'} = -3\tilde{i'} + 3\tilde{i'} = 0.
\end{align}
Hence, finally, we can write down the contraction of spin tensors in a simple way as a sum of contractions of their constituent parts:
\bea
&S^{\lambda, \mu\nu}S'_{\lambda, \mu\nu} = 2f^\mu f'_\mu -2 w_\mu w'^\mu + 2 i^{[\mu \nu]} i'_{[\mu \nu]} + 2 i^{\langle \mu \nu \rangle} i'_{\langle \mu \nu \rangle} + 3\tilde{i}\tilde{i'} + j^{\lambda \mu \nu} j'_{\lambda \mu \nu}.
\eea

\input{hyb.bbl}
%\bibliography{hyb}

\end{document}

%% file: hyb.bbl
%apsrev4-2.bst 2019-01-14 (MD) hand-edited version of apsrev4-1.bst
%Control: key (0)
%Control: author (72) initials jnrlst
%Control: editor formatted (1) identically to author
%Control: production of article title (-1) disabled
%Control: page (0) single
%Control: year (1) truncated
%Control: production of eprint (0) enabled
%